\documentclass[12pt,a4paper]{article}
\usepackage{jheppub}  
\usepackage{amssymb} 
\usepackage{amsmath}
\usepackage{mathtools}
\usepackage{amsfonts}    
\usepackage{dsfont}
\usepackage{pdfpages}
\usepackage{verbatim}
\usepackage{tensor}
\usepackage{tikz}
\usetikzlibrary{arrows,shapes,positioning}
\usetikzlibrary{decorations.markings}
\usetikzlibrary{calc}
\tikzstyle arrowstyle=[scale=1]
\tikzstyle directed=[postaction={decorate,decoration={markings,
    mark=at position .55 with {\arrow[arrowstyle]{stealth}}}}]
\usepackage{ifthen}
\usepackage{multirow}
\usepackage{longtable}
\usepackage[vcentermath]{youngtab}
\usepackage{afterpage}
    
\newcommand*\pFqskip{8mu}
\catcode`,\active
\newcommand*\pFq{\begingroup
        \catcode`\,\active
        \def ,{\mskip\pFqskip\relax}%
        \dopFq
}
\catcode`\,12
\def\dopFq#1#2#3#4#5{%
        {}_{#1}F_{#2}\biggl[\genfrac..{0pt}{}{#3}{#4};#5\biggr]%
        \endgroup
}
    
\newcounter{x}
\newcounter{y}
\newcounter{z}
\newcommand\xaxis{210}
\newcommand\yaxis{-30}
\newcommand\zaxis{90}
\newcommand\topside[3]{
  \fill[fill=red!60!black, draw=black,shift={(\xaxis:#1)},shift={(\yaxis:#2)},
  shift={(\zaxis:#3)}] (0,0) -- (30:1) -- (0,1) --(150:1)--(0,0);
}
\newcommand\leftside[3]{
  \fill[fill=red!90!black, draw=black,shift={(\xaxis:#1)},shift={(\yaxis:#2)},
  shift={(\zaxis:#3)}] (0,0) -- (0,-1) -- (210:1) --(150:1)--(0,0);
}
\newcommand\rightside[3]{
  \fill[fill=red!30!black, draw=black,shift={(\xaxis:#1)},shift={(\yaxis:#2)},
  shift={(\zaxis:#3)}] (0,0) -- (30:1) -- (-30:1) --(0,-1)--(0,0);
}
\newcommand\cube[3]{
  \topside{#1}{#2}{#3} \leftside{#1}{#2}{#3} \rightside{#1}{#2}{#3}
}
\newcommand\planepartition[1]{
\begin{tikzpicture}[scale=.5];
\fill[red!100!black,opacity=.1] (0,0) -- (\yaxis:7) -- (30:7) -- (\zaxis:7);
\fill[red!50!black,opacity=.1] (0,0) -- (\yaxis:7) -- (270:7) -- (\xaxis:7);
\fill[red!0!black,opacity=.1] (0,0) -- (\xaxis:7) -- (150:7) -- (\zaxis:7);
\draw[thick,->] (0,0) -- (\xaxis:7.5) node[below] {$x_1$};
\draw[thick,->] (0,0) -- (\yaxis:7.5) node[below] {$x_2$};
\draw[thick,->] (0,0) -- (\zaxis:7.5) node[right] {$x_3$};
\foreach \a in {1,...,7} {
\draw[ultra thin, dashed] (\yaxis:{\a}) -- ++(0,7);
}
\foreach \a in {1,...,7} {
\draw[ultra thin, dashed] (\xaxis:{\a}) -- ++(0,7);
}
\foreach \a in {1,...,7} {
\draw[ultra thin, dashed] (\yaxis:{\a}) -- ++(-6.06,-3.5);
}
\foreach \a in {1,...,7} {
\draw[ultra thin, dashed] (\xaxis:{\a}) -- ++(6.06,-3.5);
}
\foreach \a in {1,...,7} {
\draw[ultra thin, dashed] (\zaxis:{\a}) -- ++(-6.06,-3.5);
}
\foreach \a in {1,...,7} {
\draw[ultra thin, dashed] (\zaxis:{\a}) -- ++(6.06,-3.5);
}
 \setcounter{x}{-1}
  \foreach \a in {#1} {
    \addtocounter{x}{1}
    \setcounter{y}{-1}
    \foreach \b in \a {
      \addtocounter{y}{1}
      \setcounter{z}{-1}
      \foreach \c in {0,...,\b} {
        \addtocounter{z}{1}
      \ifthenelse{\c=0}{\setcounter{z}{-1},\addtocounter{y}{0}}{
        \cube{\value{x}}{\value{y}}{\value{z}}}
      }
    }
  }
\end{tikzpicture}  
}
\newcommand{\ba}{\begin{align}}

\newcommand{\be}{\begin{equation}}
\newcommand{\ee}{\end{equation}}

\newcommand{\ket}[1]{| #1 \rangle}
\newcommand*{\tran}{^{\mkern-1.5mu\mathsf{T}}}

\newcommand{\gluingtwo}[7]{ 
\begin{tikzpicture}[baseline={([yshift=-.5ex]current bounding box.center)}]
\draw[thick] (0,-1) -- (0,.5);
\draw[thick] (0,0) -- (.5,0);
\draw[thick] (0,-.5) -- (.5,-.5);
\node at (-.25,-.25) {#2};
\node at (.25,.25) {#3};
\node at (.25,-.25) {#4};
\node at (.25,-.75) {#5};
\node at (.25,-1.25) {#6};
\node at (.25,-1.75) {spin #7};
\ifnum \numexpr(#2+1)*(#3+1)*(#4+1)-#1=0
 \fill[red] (0,0) circle (.1);
\fi
\ifnum \numexpr(#2+1)*(#4+1)*(#5+1)-#1=0
 \fill[red] (0,-.5) circle (.1);
\fi
\end{tikzpicture}
}

\newcommand{\gluingthree}[8]{ 
\begin{tikzpicture}[baseline={([yshift=-.5ex]current bounding box.center)}]
\draw[thick] (0,-1.5) -- (0,.5);
\draw[thick] (0,0) -- (.5,0);
\draw[thick] (0,-.5) -- (.5,-.5);
\draw[thick] (0,-1) -- (.5,-1);
\node at (-.25,-.5) {#2};
\node at (.25,.25) {#3};
\node at (.25,-.25) {#4};
\node at (.25,-.75) {#5};
\node at (.25,-1.25) {#6};
\node at (.25,-1.75) {#7};
\node at (.25,-2.25) {spin #8};
\ifnum \numexpr(#2+1)*(#3+1)*(#4+1)-#1=0
 \fill[red] (0,0) circle (.1);
\fi
\ifnum \numexpr(#2+1)*(#4+1)*(#5+1)-#1=0
 \fill[red] (0,-.5) circle (.1);
\fi
\ifnum \numexpr(#2+1)*(#5+1)*(#6+1)-#1=0
 \fill[red] (0,-1) circle (.1);
\fi
\end{tikzpicture}
}

\newcommand{\gluingfour}[9]{ 
\begin{tikzpicture}[baseline={([yshift=-.5ex]current bounding box.center)}]
\draw[thick] (0,-2) -- (0,.5);
\draw[thick] (0,0) -- (.5,0);
\draw[thick] (0,-.5) -- (.5,-.5);
\draw[thick] (0,-1) -- (.5,-1);
\draw[thick] (0,-1.5) -- (.5,-1.5);
\node at (-.25,-.75) {#2};
\node at (.25,.25) {#3};
\node at (.25,-.25) {#4};
\node at (.25,-.75) {#5};
\node at (.25,-1.25) {#6};
\node at (.25,-1.75) {#7};
\node at (.25,-2.25) {#8};
\node at (.25,-2.75) {spin #9};
\ifnum \numexpr(#2+1)*(#3+1)*(#4+1)-#1=0
 \fill[red] (0,0) circle (.1);
\fi
\ifnum \numexpr(#2+1)*(#4+1)*(#5+1)-#1=0
 \fill[red] (0,-.5) circle (.1);
\fi
\ifnum \numexpr(#2+1)*(#5+1)*(#6+1)-#1=0
 \fill[red] (0,-1) circle (.1);
\fi
\ifnum \numexpr(#2+1)*(#6+1)*(#7+1)-#1=0
 \fill[red] (0,-1.5) circle (.1);
\fi
\end{tikzpicture}
}

\newcommand{\gluingfive}[9]{ 
\begin{tikzpicture}[baseline={([yshift=-.5ex]current bounding box.center)}]
\pgfmathtruncatemacro{\ten}{2*(#7)-#6} 
\draw[thick] (0,-2.5) -- (0,.5);
\draw[thick] (0,0) -- (.5,0);
\draw[thick] (0,-.5) -- (.5,-.5);
\draw[thick] (0,-1) -- (.5,-1);
\draw[thick] (0,-1.5) -- (.5,-1.5);
\draw[thick] (0,-2) -- (.5,-2);
\node at (-.25,-1) {#2};
\node at (.25,.25) {#3};
\node at (.25,-.25) {#4};
\node at (.25,-.75) {#5};
\node at (.25,-1.25) {#6};
\node at (.25,-1.75) {#7};
\node at (.25,-2.25) {\ten};
\node at (.25,-2.75) {#8};
\node at (.25,-3.25) {spin #9};
\ifnum \numexpr(#2+1)*(#3+1)*(#4+1)-#1=0
 \fill[red] (0,0) circle (.1);
\fi
\ifnum \numexpr(#2+1)*(#4+1)*(#5+1)-#1=0
 \fill[red] (0,-.5) circle (.1);
\fi
\ifnum \numexpr(#2+1)*(#5+1)*(#6+1)-#1=0
 \fill[red] (0,-1) circle (.1);
\fi
\ifnum \numexpr(#2+1)*(#6+1)*(#7+1)-#1=0
 \fill[red] (0,-1.5) circle (.1);
\fi
\ifnum \numexpr(#2+1)*(#7+1)*(2*(#7)-#6+1)-#1=0
 \fill[red] (0,-2) circle (.1);
\fi
\end{tikzpicture}
}

\newcommand{\gluingtwobare}[5]{ 
\begin{tikzpicture}[baseline={([yshift=-.5ex]current bounding box.center)}]
\draw[thick] (0,-1) -- (0,.5);
\draw[thick] (0,0) -- (.5,0);
\draw[thick] (0,-.5) -- (.5,-.5);
\node at (-.25,-.25) {#2};
\node at (.25,.25) {#3};
\node at (.25,-.25) {#4};
\node at (.25,-.75) {#5};
\ifnum \numexpr(#2+1)*(#3+1)*(#4+1)-#1=0
 \fill[red] (0,0) circle (.1);
\fi
\ifnum \numexpr(#2+1)*(#4+1)*(#5+1)-#1=0
 \fill[red] (0,-.5) circle (.1);
\fi
\end{tikzpicture}
}

\newcommand{\gluingthreebare}[6]{ 
\begin{tikzpicture}[baseline={([yshift=-.5ex]current bounding box.center)}]
\draw[thick] (0,-1.5) -- (0,.5);
\draw[thick] (0,0) -- (.5,0);
\draw[thick] (0,-.5) -- (.5,-.5);
\draw[thick] (0,-1) -- (.5,-1);
\node at (-.25,-.5) {#2};
\node at (.25,.25) {#3};
\node at (.25,-.25) {#4};
\node at (.25,-.75) {#5};
\node at (.25,-1.25) {#6};
\ifnum \numexpr(#2+1)*(#3+1)*(#4+1)-#1=0
 \fill[red] (0,0) circle (.1);
\fi
\ifnum \numexpr(#2+1)*(#4+1)*(#5+1)-#1=0
 \fill[red] (0,-.5) circle (.1);
\fi
\ifnum \numexpr(#2+1)*(#5+1)*(#6+1)-#1=0
 \fill[red] (0,-1) circle (.1);
\fi
\end{tikzpicture}
}

\newcommand{\gluingfourbare}[7]{ 
\begin{tikzpicture}[baseline={([yshift=-.5ex]current bounding box.center)}]
\draw[thick] (0,-2) -- (0,.5);
\draw[thick] (0,0) -- (.5,0);
\draw[thick] (0,-.5) -- (.5,-.5);
\draw[thick] (0,-1) -- (.5,-1);
\draw[thick] (0,-1.5) -- (.5,-1.5);
\node at (-.25,-.75) {#2};
\node at (.25,.25) {#3};
\node at (.25,-.25) {#4};
\node at (.25,-.75) {#5};
\node at (.25,-1.25) {#6};
\node at (.25,-1.75) {#7};
\ifnum \numexpr(#2+1)*(#3+1)*(#4+1)-#1=0
 \fill[red] (0,0) circle (.1);
\fi
\ifnum \numexpr(#2+1)*(#4+1)*(#5+1)-#1=0
 \fill[red] (0,-.5) circle (.1);
\fi
\ifnum \numexpr(#2+1)*(#5+1)*(#6+1)-#1=0
 \fill[red] (0,-1) circle (.1);
\fi
\ifnum \numexpr(#2+1)*(#6+1)*(#7+1)-#1=0
 \fill[red] (0,-1.5) circle (.1);
\fi
\end{tikzpicture}
}

\newcommand{\gluingfivebare}[7]{ 
\begin{tikzpicture}[baseline={([yshift=-.5ex]current bounding box.center)}]
\pgfmathtruncatemacro{\ten}{2*(#7)-#6} 
\draw[thick] (0,-2.5) -- (0,.5);
\draw[thick] (0,0) -- (.5,0);
\draw[thick] (0,-.5) -- (.5,-.5);
\draw[thick] (0,-1) -- (.5,-1);
\draw[thick] (0,-1.5) -- (.5,-1.5);
\draw[thick] (0,-2) -- (.5,-2);
\node at (-.25,-1) {#2};
\node at (.25,.25) {#3};
\node at (.25,-.25) {#4};
\node at (.25,-.75) {#5};
\node at (.25,-1.25) {#6};
\node at (.25,-1.75) {#7};
\node at (.25,-2.25) {\ten};
\ifnum \numexpr(#2+1)*(#3+1)*(#4+1)-#1=0
 \fill[red] (0,0) circle (.1);
\fi
\ifnum \numexpr(#2+1)*(#4+1)*(#5+1)-#1=0
 \fill[red] (0,-.5) circle (.1);
\fi
\ifnum \numexpr(#2+1)*(#5+1)*(#6+1)-#1=0
 \fill[red] (0,-1) circle (.1);
\fi
\ifnum \numexpr(#2+1)*(#6+1)*(#7+1)-#1=0
 \fill[red] (0,-1.5) circle (.1);
\fi
\ifnum \numexpr(#2+1)*(#7+1)*(2*(#7)-#6+1)-#1=0
 \fill[red] (0,-2) circle (.1);
\fi
\end{tikzpicture}
}

\allowdisplaybreaks[1]

\title{The matrix-extended $\boldsymbol{\mathcal{W}_{1+\infty}}$ algebra}

\author[a,b]{Lorenz Eberhardt}
\author[c]{Tom\'{a}\v{s} Proch\'{a}zka,}
\affiliation[a]{Institut f\"ur Theoretische Physik, ETH Zurich, \\
\hspace*{0.3cm}CH-8093 Z\"urich, Switzerland}
\affiliation[b]{School of Natural Sciences, Institute for Advanced Study, \\
\hspace*{0.3cm}Princeton, NJ 08540, USA}
\affiliation[c]{Arnold Sommerfeld Center for Theoretical Physics, \\
\hspace*{0.3cm}Ludwig Maximilian University of Munich, \\
\hspace*{0.3cm}Theresienstr.~37, D-80333 M\"unchen, Germany}
\emailAdd{elorenz@ias.edu}
\emailAdd{Tomas.Prochazka@lmu.de}

\abstract{We construct a quadratic basis of generators of matrix-extended $\mathcal{W}_{1+\infty}$ using a generalization of the Miura transformation. This makes it possible to conjecture a closed-form formula for the operator product expansions defining the algebra. We study truncations of the algebra. An explicit calculation at low levels shows that these are parametrized in a way consistent with the gluing description of the algebra. It is perhaps surprising that in spite of the fact that the algebras are rather complicated and non-linear, the structure of their truncations follows very simple gluing rules.}

\begin{document}
\maketitle

\makeatletter
\g@addto@macro\bfseries{\boldmath}
\makeatother
\section{Introduction}
$\mathcal{W}_\infty$ algebras have attracted renewed interest in the last years. They are some of the most basic symmetry algebras that arise in 2d CFTs. The algebra $\mathcal{W}_N$ is the unique chiral algebra that has besides the stress tensor a field of every spin $s\in \{3,4,\dots,N\}$ \cite{Zamolodchikov:1985wn, Bouwknegt:1992wg}.
$\mathcal{W}_N$ algebras arise as the symmetry algebra in the AGT correspondence \cite{Alday:2009aq, Wyllard:2009hg}, which relates 4d $\mathcal{N}=2$ $\mathrm{SU}(N)$ gauge theory with Toda theory. Toda theory is the natural $\mathcal{W}$-analogue of Liouville theory. However, unlike in Liouville theory, a closed-form expression for the three-point functions of Toda theory still seems out of reach. $\mathcal{W}_N$ algebras also arise naturally in the 4d $\mathcal{N=2}$ superconformal theories and 6d $\mathcal{N}=(2,0)$ theory, where they appear as the cohomology of a certain superconformal supercharge \cite{Beem:2013sza,Beem:2014kka}.

$\mathcal{W}_N$ algebras have played a central role in the higher spin $\mathrm{AdS}_3/\mathrm{CFT}_2$ correspondence \cite{Gaberdiel:2011wb}. The matching of this symmetry algebra in the correspondence was one of the main pieces of evidences for the duality.

All $\mathcal{W}_N$ algebras stem from the $\mathcal{W}_\infty[c,\lambda]$ algebra, which depends beyond the central charge $c$ also on a further parameter $\lambda$ \cite{Pope:1989ew, Pope:1989sr, Pope:1990kc, Gaberdiel:2011wb, Gaberdiel:2012ku, Prochazka:2014gqa, Linshaw:2017tvv}. This parameter is in a sense the analytic continuation of $N$. $\mathcal{W}_\infty[c,\lambda]$ has a generating field of every spin $s\in \mathds{Z}_{\ge 2}$. If $\lambda=N\in \mathds{Z}_{\ge 2}$, the algebra develops an ideal and can be consistently truncated to $\mathcal{W}_N$. Hence $\mathcal{W}_\infty[c,\lambda]$ unifies all $\mathcal{W}_N$ algebras. It has a much richer structure than $\mathcal{W}_N$, since it depends on two parameters $c$ and $\lambda$. It was realized in \cite{Gaberdiel:2012ku} that there are in fact three values of $\lambda$ for which the resulting algebra is isomorphic:
\be 
\mathcal{W}_\infty[c,\lambda_1]\cong \mathcal{W}_\infty[c,\lambda_2]\cong \mathcal{W}_\infty[c,\lambda_3]\ .
\ee 
This phenomenon is called triality, which acts as an outer automorphism of the algebra.

For many purposes, it is useful to add a free $\mathfrak{u}(1)$ current to the algebra and consider the algebra $\mathcal{W}_{1+\infty}[c,\lambda]\cong \mathfrak{u}(1)\times\mathcal{W}_\infty[c,\lambda]$, which has one generating higher spin field of every spin $s \in \mathds{N}$.
This algebra is isomorphic to the affine Yangian of $\mathfrak{gl}(1)$ \cite{Tsymbaliuk:2014fvq, Prochazka:2015deb, Negut:2016dxr, Gaberdiel:2017dbk}, and this connection implies a remarkable properties of the representation theory for $\mathcal{W}_{1+\infty}[c,\lambda]$. The representation theory of the algebra can be formulated in terms of plane partitions, i.e.~3d partitions. Via this correspondence, states in the vacuum representation of $\mathcal{W}_{1+\infty}$ can be labeled by plane partitions and so their generating function -- the MacMahon function -- is the vacuum character of $\mathcal{W}_{1+\infty}$. This picture also generalizes to non-trivial representations of the algebra.
They correspond to plane-partitions with non-trivial asymptotics along the three coordinate axes. Thus, representations are in general labeled by three 2d Young diagrams which specify the asymptotics of the 3d partitions. An example of such a configuration is drawn in Figure~\ref{fig:example state}.
The triality symmetry is manifest in the Yangian description, it simply permutes the three coordinate axes of the plane-partitions.
\begin{figure}
\begin{center}
\planepartition{{7,3,2,2,2,2,2},{4,2,2,1,1,1,1},{2,2},{2,2},{2,2},{2,2},{2,2}}
\end{center}
\caption{A plane partition in the representation specified by the three Young diagrams ${\tiny \protect\yng(2,2)}$, ${\tiny \protect\yng(2,1)}$ and ${\tiny \protect\yng(1)}$.} \label{fig:example state}
\end{figure}

In two-dimensional CFTs, one is often interested in truncations of the algebra for special values of parameters. For instance the codimension $2$ truncations of $\mathcal{W}_{1+\infty}$ lead to rational models. In the context of $\mathcal{W}_{1+\infty}$, codimension $1$ truncations are well-understood. They are labeled by triples of non-negative integers $(N_1,N_2,N_3)$. Such a triple specifies a one-dimensional curve in the parameter space $\lambda$ where the algebra truncates. These algebras were originally constructed using a brane-construction and they arise at the triple $Y$-shaped junction of supersymmetric interfaces in $\mathcal{N}=4$ SYM. For this reason, they were given the name $Y_{N_1,N_2,N_3}$ algebra \cite{Gaiotto:2017euk, Prochazka:2017qum}. In the Yangian picture, these truncations have again a very simple interpretation -- they correspond to constrained plane partitions in which the box at position $(N_1,N_2,N_3)$ is not allowed.
Minimal model CFTs correspond to intersections of two such truncation curves and in the Yangian picture to periodic plane partitions. This gives in particular a very simple unifying combinatorial characterization of the characters of these models.

Given the success of this program for the $\mathcal{W}_{1+\infty}$ algebra, one may ask whether there is an analogous construction for different or larger algebras. In particular, one would like to capture other (quasi-)rational minimal models such as the $\mathfrak{su}(2)_k$ WZW model at rational level $k=-2+\frac{p}{q}$ with $p\ge 2$ and $\mathrm{gcd}(p,q)=1$ or supersymmetric algebras. 
One example of such an extension is the $\mathcal{N}=2$ $\mathcal{W}_{1+\infty}$ algebra, to which a supersymmetric version of the affine Yangian \cite{Gaberdiel:2017hcn, Prochazka:2017qum,Gaberdiel:2018nbs} can be associated. 

In this work, we make further progress towards achieving this goal. We do so by considering a flavored extension of $\mathcal{W}_{1+\infty}$, i.e.~we add matrix degrees of freedom. The algebra we want to consider possesses $m^2$ generating fields of every spin $s \in \mathds{N}$. In particular, the spin 1 fields define an affine Kac-Moody algebra $\mathfrak{gl}(m)_k$ and the higher spin fields transform in the adjoint representation of this Kac-Moody algebra. It turns out that there is unique algebra (depending now on three parameters -- the central charge $c$ and the analogous parameter to $\lambda$ as well as the matrix rank $m$) with this spin content. This algebra and its cousins was studied before in \cite{Costello:2017fbo, Eberhardt:2018plx, Creutzig:2018pts, Creutzig:2019qos, Creutzig:2019wfe}, where it was sometimes referred to as `rectangular $\mathcal{W}$-algebra'. Moreover, its geometric realizations are described in \cite{Rapcak:toappear}.

We find that these algebras have in fact a very beautiful structure. Analogously to the situation in $\mathcal{W}_{1+\infty}$ \cite{Lukyanov:1988aa,Prochazka:2014gqa}, there seems to be a basis of generators in which the algebra becomes especially simple. We construct this basis naturally by defining a matrix-extended Miura transformation. In this extended Miura transformation, we start with $n$ copies of Kac-Moody algebra $\mathfrak{gl}(m)_\kappa$ and construct a matrix-valued differential operator. This defines matrix-valued higher spin fields defining the matrix-extended $\mathcal{W}_{1+\infty}$ algebra. In this basis, the non-linearity of the algebra is only quadratic. We also find that the structure constants are independent of $m$. Thus the matrix-structure is straightforward to incorporate. Since the algebra simplifies so significantly, we are able to guess a closed-form formula for the structure constants of the algebra.

The matrix-extension of the algebra breaks the triality symmetry of $\mathcal{W}_{1+\infty}$ down to a duality symmetry. This is however partially compensated by the fact that the matrix-structure leads to the appearance of other outer automorphism generators. These are given by spectral flow, which extends to the whole matrix-extended algebra. Thus the full outer automorphism group (for $m \ge 3$) turns out to be $\mathds{Z}_2 \times \mathrm{D}_m$, where $\mathrm{D}_m$ is the dihedral group with $m$ elements.

We then study truncations of the algebra and find that they follow a very intuitive pattern. Analogously to the $m=1$ case, truncations are specified by $m+2$ integers, $4$ of which are independent.

This paper is organized as follows. In Section~\ref{sec:review}, we review the most important features of $\mathcal{W}_{1+\infty}$ and the Miura transformation. We discuss the matrix-extended Miura transformation and the structure of the algebra in Section~\ref{sec:structure}. Section~\ref{sec:representations} explores the representation theory and truncations of the algebra. We end with a discussion and outlook in Section~\ref{sec:discussion}.

\section{Review of the \texorpdfstring{$\boldsymbol{\mathcal{W}_{1+\infty}}$}{W1infty} algebra} \label{sec:review}
Before initiating our study of the matrix-extended $\mathcal{W}_{1+\infty}$ algebra, we review the most salient features of the $\mathcal{W}_{1+\infty}$ algebra.

\subsection{The Miura transformation}\label{subsec:Miura review}
The algebra $\widehat{\mathfrak{u}}(1) \times \mathcal{W}_n$ can be defined as follows. Let $J$ be a $\widehat{\mathfrak{u}}(1)$ current with defining OPE
\be 
J(z) J(w) \sim \frac{\kappa+1}{(z-w)^2}\ , \label{eq:u(1) ope}
\ee
where $\kappa \in \mathds{C}\setminus \{-1\}$ is a complex parameter. This choice of normalization will be very useful in what follows. Consider now the differential Miura operator
\be 
\mathcal{L}(z)=\prod_{i=1}^n (-\kappa\, \partial+J^{(i)}(z))\ ,
\ee
where we employed $n$ copies of the $\mathrm{U}(1)$ current. The ordering of the product is such that the index $(i)$ increases from left to right (for discussion of other orderings and the associated $R$-matrix see \cite{Prochazka:2019dvu}). We can rewrite
\be 
\mathcal{L}(z)=\sum_{k=0}^n U_{(k)}(z) (-\kappa\, \partial)^{n-k}\ ,
\ee
thereby defining the spin $k$ fields $U_{(k)}(z)$. $U_{(k)}$ read \cite{Prochazka:2014gqa}
\begin{subequations}
\begin{align}
U_{(1)}&=\sum_{i=1}^n J^{(i)}\ ,\\
U_{(2)}&=\sum_{i<j}^n (J^{(i)}J^{(j)})-\kappa \sum_{i=1}^n (i-1) \partial J^{(i)}\ , \\
U_{(3)}&=\sum_{i<j<k}^n (J^{(i)}J^{(j)}J^{(k)})-\kappa \sum_{i<j}^n \left((i-1) \partial J^{(i)} J^{(j)}+(j-2)J^{(i)} \partial J^{(j)}\right)\nonumber\\
&\qquad+\frac{\kappa^2}{2}\sum_{i=1}^n(i-1)(i-2) \partial^2 J^{(i)}\ , \\
U_{(k)}&=\sum_{r=1}^k (-\kappa)^{k-r} \sum_{\genfrac{}{}{0pt}{}{i_1<\cdots<i_r}{\ell_1+\cdots+\ell_r=k-r}} \prod_{j=1}^r \binom{i_j-\sum_{s=1}^{j-1}\ell_s-j}{\ell_j} \partial^{\ell_j} J^{(i_j)}\ ,
\end{align}
\end{subequations}
where we wrote down the first three fields explicitly.
It is a well-known fact \cite{Fateev:1987zh, Lukyanov:1988aa, Bouwknegt:1992wg} that the OPEs of these fields close on themselves and are strong generators for the algebra $\widehat{\mathfrak{u}}(1)\times \mathcal{W}_n$.

Moreover, this basis (to which we shall henceforth refer to as Miura or quadratic basis) of $\mathcal{W}_n$ has several advantages. While the fields $U_{(k)}$ are not quasi-primary, the OPEs have at most quadratic non-linearity in this basis and hence take the form \cite{Lukyanov:1988aa, Prochazka:2014gqa}
\begin{align}
U_{(j)}(z) U_{(k)}(w) \sim \sum_{\ell,m,r,s} C_{j,k}^{\ell,m,r,s}(\kappa,n)\, \frac{(\partial^r U_{(\ell)} \partial^s U_{(m)})(w)}{(z-w)^{j+k-r-s-\ell-m}}\ ,
\end{align}
where $C_{j,k}^{\ell,m,r,s}(\kappa,n)$ are the structure constants of the algebra, which are polynomial in this basis. This structure simplies further. It was noticed in \cite{Prochazka:2014gqa} that one can rewrite the OPE in the following form
\be 
U_{(j)}(z)U_{(k)}(w)\sim-\sum_{\ell+m<j+k} D_{j,k}^{\ell,m}(\kappa,n)\frac{U_{(\ell)}(z)U_{(m)}(w)}{(z-w)^{j+k-\ell-m}}\ ,
\ee
where the coefficients $D_{j,k}^{\ell,m}(\kappa,n)$ are also polynomial in $n$ and $\kappa$. Notice that this fixes the entire structure of the derivatives recursively. Hence, knowing only $C_{j,k}^{\ell,m,0,0}(\kappa,n)$, one can reconstruct the entire OPE including derivatives. From now on, we will write $C_{j,k}^{\ell,m}(\kappa,n)\equiv C_{j,k}^{\ell,m,0,0}(\kappa,n)$. Moreover, it was observed in \cite{Prochazka:2014gqa} that the structure constants obey a `shift symmetry' of the form
\be 
C_{j,k}^{\ell,m}(\kappa,n)=C_{j-1,k-1}^{\ell-1,m-1}(\kappa,n-1)\ ,
\ee
and thus one can always reduce to the case where $m=0$ (in the case that $\ell=0$ instead, we can use the fact that $C_{j,k}^{\ell, m}(\kappa,n) = (-1)^{j+k-\ell-m} C_{k,j}^{m, \ell}(\kappa,n)$).\footnote{It could also happen that $j=0$ or $k=0$. But for $j=0$, we have clearly $C_{0,k}^{\ell,m}(\kappa,n)=\delta^\ell_0 \delta^m_k$. We will exclude this trivial case.} The result for the structure constants can be found in \cite{Prochazka:2014gqa}.

It was discussed extensively in the literature \cite{Gaberdiel:2011wb, Prochazka:2014gqa}, that $\mathcal{W}_n$ can be realized as a truncation of a bigger algebra, the so-called $\mathcal{W}_{\infty}[\lambda]$ algebra, which contains one strong generator for every spin $s \in \mathds{N}_{\ge 2}$. In the following, we will often consider the algebra $\mathcal{W}_{1+\infty}[\lambda] \equiv \widehat{\mathfrak{u}}(1) \times \mathcal{W}_\infty[\lambda]$, since many properties are more conveniently expressed in this formulation. $\mathcal{W}_{1+\infty}[\lambda]$ depends besides the central charge on additional parameter $\lambda$. Upon specialization of $\lambda=n\in \mathds{N}$, the algebra forms an ideal and the relevant truncation is again $\widehat{\mathfrak{u}}(1)\times \mathcal{W}_{n}$. The structure constants of $\mathcal{W}_{1+\infty}[\lambda]$ are those given in \cite{Prochazka:2014gqa}, where we simply allow $n(=\lambda)$ to take arbitrary complex values. 

A useful parametrization of the algebra is given by setting
\be 
\lambda_1=n\ , \qquad \lambda_2=-\frac{n\kappa}{\kappa+1}\ , \qquad \lambda_3=n \kappa\ , \label{eq:lambdankappa translation}
\ee
which satisfy
\be 
\label{eq:lambda sum constraint}
\frac{1}{\lambda_1}+\frac{1}{\lambda_2}+\frac{1}{\lambda_3}=0\ .
\ee

It was first noticed in \cite{Gaberdiel:2012ku} that $\mathcal{W}_{1+\infty}[\lambda]$ has a discrete triality symmetry which acts by permutation on the parameters $\lambda_i$. 

\subsection{Truncations and the gluing construction}\label{subsec:truncations and gluing}
We have already explained that $\mathcal{W}_{1+\infty}$ admits a truncation for $\lambda_1 \in \mathds{N}$ to $\widehat{\mathfrak{u}}(1)\times \mathcal{W}_n$. In \cite{Prochazka:2014gqa, Gaiotto:2017euk, Prochazka:2017qum}, it was discovered that there exist more general truncations of $\mathcal{W}_{1+\infty}$, the so-called Y-algebras.
Y-algebras are parametrized by three integers $N_1$, $N_2$ and $N_3 \in \mathds{N}_0$ and are located on curves in the two-dimensional parameter space spanned by the parameters $\lambda_i$, where
\be 
\frac{N_1}{\lambda_1}+\frac{N_2}{\lambda_2}+\frac{N_3}{\lambda_3}=1\ . \label{eq:truncation condition}
\ee
They also depend on the one remaining continuous parameter (which in the literature is conventionally denoted by $\Psi$). Note that due to relation (\ref{eq:lambda sum constraint}) the triples $N_j$ differing by an overall constant shift lead to same restriction on the parameters of the algebra.

For an arbitrary choice of parameters $N_j \geq 0$ the algebra $Y_{N_1,N_2,N_3}$ can be defined by quotienting out $\mathcal{W}_{1+\infty}$ by an ideal generated by a singular vector at level
\be 
(N_1+1)(N_2+1)(N_3+1)\ , \label{eq:null-vector level}
\ee
which corresponds to a rectangular configuration of boxes of dimension $N_1 \times N_2 \times N_3$ in the Yangian picture \cite{Gaiotto:2017euk}. Alternatively, one may write a free field representation in terms of $N_1+N_2+N_3$ free fields where this singular vector is identically equal to zero \cite{Prochazka:2018tlo}. Note that only in the case that (at least) one of $N_j$ parameters is equal to zero (and parameter $\Psi$ is generic), this quotient is simple. Since it is possible for a simple algebra to have non-simple subalgebras, in some situations (like when we consider gluing) we have to consider these more general non-simple quotients as well.\footnote{The simplest example is the algebra $Y_{111}$ with central charge $c=0$. All the fields different from the identity form an ideal and the corresponding simple quotient has just the identity operator. We can however find a free field realization of this algebra in terms of three free bosons and in this representation the first null field is at level $8$ corresponding to a rectangular configuration of $2 \times 2 \times 2$ boxes \cite{Prochazka:2018tlo}.} The simplest special case of $\widehat{\mathfrak{u}}(1)\times\mathcal{W}_{n}$ corresponds to the truncation $N_1=n$, $N_2=N_3=0$.

Y-algebras serve as fundamental building blocks for more general $\mathcal{W}$-algebras. Constructing these more complicated $\mathcal{W}$-algebras is achieved via the gluing construction \cite{Prochazka:2017qum}. Analogously to the topological vertex, we represent a $Y_{N_1,N_2,N_3}$-algebra by the vertex
\be 
\begin{tikzpicture}[baseline={([yshift=-.5ex]current bounding box.center)}]
\draw[thick,directed] (0,0) -- (1.5,0);
\draw[thick,directed] (0,0) -- (0,1.5);
\draw[thick,directed] (0,0) -- (-1.06,-1.06);
\node at (.7,.7) {$N_1$};
\node at (.5,-.7) {$N_2$};
\node at (-.7,.5) {$N_3$};
\end{tikzpicture}\ .
\ee
This is motivated by the fact that $Y_{N_1,N_2,N_3}$ algebras arise as the operator algebra of local operators living on the trivalent junction of $N_1$ D5-branes, $N_2$ NS5-branes and $N_3$ $(1,1)$-branes. When gluing two of these building blocks together, the resulting algebra depends on the relative orientation of the corresponding trivalent vertices. The change of orientation of a vertex corresponds to an action of S-duality on the branes. We assign $(p,q)$-charges to the legs of each vertex. These charges must satisfy the charge conservation condition and additional condition
\be 
\sum_i p_i=0\ ,\qquad \sum_i q_i=0\ ,\qquad p_1 q_2-p_2 q_1=1\ . \label{eq:gluing conditions}
\ee
The S-duality can be compensated by performing an appropriate change in the parameters $\lambda_i$. To state this change, it is simplest to use yet a different parametrization of the $Y_{N_1,N_2,N_3}$ algebra in terms of $\varepsilon_i$, $i=1,2,3$, which satisfy
\be 
\sum_i \varepsilon_i=0\ , \qquad \lambda_i \varepsilon_i=N_1 \varepsilon_1+N_2 \varepsilon_2+N_3 \varepsilon_3\ .
\ee
These parameters are only defined up to an overall multiple, so actually one conventionally parametrizes them in terms of $\Psi = -\frac{\epsilon_2}{\epsilon_1}$ \cite{Gaiotto:2017euk}. The $\varepsilon$-parameters of the S-dualized Y-algebra read then
\be 
\varepsilon_j=p_j \varepsilon_1+q_j \varepsilon_2\ .
\ee

The gluing construction is a way to compose several vertices to build bigger VOAs. For this, we draw diagrams like
\be 
\begin{tikzpicture}[baseline={([yshift=-.5ex]current bounding box.center)}]
\draw[thick,directed] (-1.06,1.06) -- (0,0);
\draw[thick,directed] (-1.06,-1.06) -- (0,0);
\draw[thick,directed] (1.5,0) -- (0,0);
\draw[thick,directed] (1.5,0) -- (2.56,1.06);
\draw[thick,directed] (1.5,0) -- (2.56,-1.06);
\node at (.75,.7) {$N_3$};
\node at (-.7,0) {$N_1$};
\node at (.75,-.7) {$N_2$};
\node at (2.2,0) {$N_4$};
\end{tikzpicture}\ ,
\ee
which represents a conformal extension of the algebra
\be 
Y_{N_1,N_2,N_3} \times Y_{N_3,N_2,N_4}
\ee
with the appropriate $\lambda$ parameters. The fields extending the algebra are called gluing fields and from the brane point of view correspond to line operators stretched along the edges of the diagram. For more details see \cite{Prochazka:2017qum}.

\section{The matrix-extended \texorpdfstring{$\boldsymbol{\mathcal{W}_{1+\infty}}$}{W1infty} algebra} \label{sec:structure}

In this paper, we study a matrix-extended version of $\mathcal{W}_{1+\infty}[\lambda]$. This algebra was studied previously in \cite{Costello:2017fbo, Eberhardt:2018plx, Creutzig:2018pts} and generalizations were considered in \cite{Creutzig:2019qos, Creutzig:2019wfe}. 
The purpose of this paper is to expose the structure of this algebra and to study its representation theory. The matrix structure turns out to be surprisingly simple if looked at in the correct variables and many of the properties of $\mathcal{W}_{1+\infty}[\lambda]$ still hold.

The algebra possesses $m^2$ generating fields of every spin $s \in \mathds{N}$, which transform in the adjoint representation of the global symmetry $\mathfrak{gl}(m)$. Thus, the algebra enjoys in particular a $\widehat{\mathfrak{gl}}(m)$ Kac-Moody symmetry. The algebra contains a central $\widehat{\mathfrak{u}}(1)$ current and decoupling it would lead to $\widehat{\mathfrak{sl}}(m)$ Kac-Moody symmetry. This is analogous to the decoupling of the $\widehat{\mathfrak{u}}(1)$ current in $\mathcal{W}_{1+\infty}[\lambda]$, which yields $\mathcal{W}_\infty[\lambda]$. Many of the properties of the algebra are clearest (and the description is more uniform) when keeping it.

The algebra is parametrized in terms of three parameters, which we take to be $\kappa$, $n$ and $m$. In terms of these parameters, the spin 1 fields lead to the Kac-Moody algebra
\be 
\mathfrak{gl}(m)_{n \kappa}\ ,
\ee
and the central charge of the whole algebra takes the form
\be 
c=\frac{mn}{m+\kappa}\left[1+m \kappa-(n^2-1)\kappa^2\right]\ . \label{eq:central charge}
\ee

The triality symmetry of $\mathcal{W}_{1+\infty}$ is broken to a duality symmetry in the matrix-extended case. The basic reason is that the level of the Kac-Moody algebra cannot be rescaled if $m \ge 2$ and thus any discrete symmetry has to fix the level $n \kappa$, as well as the central charge (and the matrix rank $m$). The unique such transformation is
\be 
\kappa \to -m-\kappa\ , \qquad n \to -\frac{n \kappa}{m+\kappa}\ , \label{eq:duality symmetry}
\ee
which generates the duality symmetry in this case.

\subsection{Miura transformation}
Analogously to the Miura transformation for $\widehat{\mathfrak{u}}(1) \times \mathcal{W}_{n-1}$, we can introduce a matrix-valued differential operator
\be 
\mathcal{L}(z)=\prod_{i=1}^n(-\kappa\,\mathds{1}_{m \times m}\partial+\tensor{J}{^{(i)}^a_b}(z)\tensor{\text{E}}{_a^b})=\sum_{k=0}^n \tensor{U}{_{(k)}^a_b}(z)\tensor{\text{E}}{_a^b} (-\kappa \,\partial)^{n-k}\ ,\label{eq:matrix Miura operator}
\ee
where $\tensor{J}{^{(i)}^a_b}(z)$ satisfies the $\mathfrak{gl}(m)_\kappa$ algebra. Here, $\tensor{\text{E}}{_a^b}$ are elementary matrices forming a basis for the Lie algebra $\mathfrak{gl}(m)$, satisfying $\tensor{\text{E}}{_a^b}\tensor{\text{E}}{_c^d}=\tensor{\delta}{^b_c}\tensor{\text{E}}{_a^d}$. Again, we chose the convention to label the fields in the product from left to right. We have
\be 
\tensor{J}{^{(i)}^a_b}(z) \tensor{J}{^{(i)}^c_d}(w) \sim \frac{\tensor{\delta}{^a_b} \tensor{\delta}{^c_d} + \kappa \, \tensor{\delta}{^c_b} \tensor{\delta}{^a_d}}{(z-w)^2} + \frac{\tensor{\delta}{^c_b} \tensor{J}{^{(i)}^a_d}(w) - \tensor{\delta}{^a_d} \tensor{J}{^{(i)}^c_b}(w)}{z-w}\ . \label{eq:glm normalization convention}
\ee
The normalization of the $\widehat{\mathfrak{u}}(1)$ current $\tensor{J}{^{(i)}^a_a}$ is important and is the natural generalization of \eqref{eq:u(1) ope}. This defines the matrix-valued higher spin fields $U_{(k)}(z)$. Explicitly, we have in components
\begin{subequations}
\begin{align}
\tensor{U}{_{(1)}^a_b}&=\sum_{i=1}^n \tensor{J}{^{(i)}^a_b}\ ,\\
\tensor{U}{_{(2)}^a_b}&=\sum_{i<j}^n \tensor{J}{^{(i)}^a_c}\tensor{J}{^{(j)}^c_b}-\kappa \sum_{i=1}^n (i-1) \partial \tensor{J}{^{(i)}^a_b}\ , \\
\tensor{U}{_{(3)}^a_b}&=\sum_{i<j<k}^n \tensor{J}{^{(i)}^a_c}\tensor{J}{^{(j)}^c_d}\tensor{J}{^{(k)}^d_b}-\kappa \sum_{i<j}^n \left((i-1) \partial \tensor{J}{^{(i)}^a_c} \tensor{J}{^{(j)}^c_b}+(j-2)\tensor{J}{^{(i)}^a_c} \partial \tensor{J}{^{(j)}^c_b}\right)\nonumber\\
&\qquad+\frac{\kappa^2}{2}\sum_{i=1}^n(i-1)(i-2) \partial^2 \tensor{J}{^{(i)}^a_b}\ .
\end{align}
\end{subequations}
which (except for the matrix multiplication and replacement $\alpha_0 \to -\kappa$) is exactly of the same form as the corresponding fields in $\mathcal{W}_{1+\infty}$ \cite{Prochazka:2014gqa}. These expressions are actually special cases of the fusion coproduct in matrix-extended $\mathcal{W}_{1+\infty}$ which (except for the additional contraction of indices) is of the same form as in \cite{Prochazka:2014gqa}. The existence of this coproduct is a direct consequence of the multiplicative structure of the Miura operators.

By explicit calculation, we have shown that the higher spin fields $U_{(k)}$ close on themselves under OPEs.\footnote{We have checked this for the OPE $\tensor{U}{_{(j)}^a_b}(z)\tensor{U}{_{(k)}^c_d}(w)$ for $j+k\le 7$. We do however not have a general proof of this fact.} In fact, this is only true if we choose the normalization of the $\widehat{\mathfrak{u}}(1)$ current in the way we did. One can always change the normalization of the generators of the diagonal subalgebra, but fixing the form of the Miura factor (\ref{eq:matrix Miura operator}) the normalization of the currents is already fixed if we want the algebra to close quadratically. Notice that the quadratic pole in (\ref{eq:glm normalization convention}) is actually an $R$-matrix of $\mathfrak{gl}(m)$ and satisfies the Yang-Baxter equation.
The low-lying OPEs take the form
\begin{subequations}
\begin{align}
\tensor{U}{_{(1)}^a_b}(z)\tensor{U}{_{(1)}^c_d}(w) &\sim\frac{n (\kappa  \tensor{\delta}{^a_d} \tensor{\delta}{^c_b}+\tensor{\delta}{^a_b} \tensor{\delta}{^c_d})}{(z-w)^2}+ \frac{\tensor{\delta}{^c_b} \tensor{U}{_{(1)}^a_d}(w)-\tensor{\delta}{^a_d} \tensor{U}{_{(1)}^c_b}(w)}{z-w}\ , \label{eq:U1U1 OPE}\\
\tensor{U}{_{(1)}^a_b}(z)\tensor{U}{_{(2)}^c_d}(w)&\sim-\frac{\kappa  (n-1) n (\kappa  \tensor{\delta}{^a_d} \tensor{\delta}{^c_b}+\tensor{\delta}{^a_b} \tensor{\delta}{^c_d})}{(z-w)^3}\nonumber\\
&\qquad+\frac{(n-1) (\kappa  \tensor{\delta}{^a_d} \tensor{U}{_{(1)}^c_b}(w)+\tensor{\delta}{^a_b} \tensor{U}{_{(1)}^c_d}(w))}{(z-w)^2}\nonumber\\
&\qquad+\frac{\tensor{\delta}{^c_b} \tensor{U}{_{(2)}^a_d}(w)-\tensor{\delta}{^a_d} \tensor{U}{_{(2)}^c_b}(w)}{z-w}\ , \label{eq:U1U2 OPE}\\
\tensor{U}{_{(1)}^a_b}(z)\tensor{U}{_{(3)}^c_d}(w) &\sim\frac{\kappa ^2 (n-2) (n-1) n (\kappa  \tensor{\delta}{^a_d} \tensor{\delta}{^c_b}+\tensor{\delta}{^a_b} \tensor{\delta}{^c_d})}{(z-w)^4}\nonumber\\
&\qquad-\frac{\kappa  (n-2) (n-1) (\kappa  \tensor{\delta}{^a_d} \tensor{U}{_{(1)}^c_b}(w)+\tensor{\delta}{^a_b} \tensor{U}{_{(1)}^c_d}(w))}{(z-w)^3}\nonumber\\
&\qquad+\frac{(n-2) (\kappa  \tensor{\delta}{^a_d} \tensor{U}{_{(2)}^c_b}(w)+\tensor{\delta}{^a_b} \tensor{U}{_{(2)}^c_d}(w))}{(z-w)^2}\nonumber\\
&\qquad+\frac{\tensor{\delta}{^c_b} \tensor{U}{_{(3)}^a_d}(w)-\tensor{\delta}{^a_d} \tensor{U}{_{(3)}^c_b}(w)}{z-w}\ ,\label{eq:U1U3 OPE} \\
\tensor{U}{_{(2)}^a_b}(z)\tensor{U}{_{(2)}^c_d}(w) &\sim\frac{n(n-1)(\kappa(1+\kappa^2-2n \kappa^2)\tensor{\delta}{^a_d}\tensor{\delta}{^c_b})+(1+3\kappa^2-4n\kappa^2)\tensor{\delta}{^a_b}\tensor{\delta}{^c_d})}{2(z-w)^4}\nonumber\\
&\qquad-\frac{(n-1)(n\kappa^2-1)(\tensor{\delta}{^c_b} \tensor{U}{_{(1)}^a_d}(w)-\tensor{\delta}{^a_d} \tensor{U}{_{(1)}^c_b}(w))}{(z-w)^3}\nonumber\\
&\qquad-\frac{(n-1)^2\kappa (\tensor{\delta}{^c_d} \tensor{U}{_{(1)}^a_b}(w)-\tensor{\delta}{^a_b} \tensor{U}{_{(1)}^c_d}(w))}{(z-w)^3}\nonumber\\
&\qquad-\frac{\tensor{\delta}{^c_d} \tensor{U}{_{(2)}^a_b}(w)+\tensor{\delta}{^a_b} \tensor{U}{_{(2)}^c_d}(w)+\kappa\tensor{\delta}{^c_b} \tensor{U}{_{(2)}^a_d}(w)+\kappa\tensor{\delta}{^a_d} \tensor{U}{_{(2)}^c_b}(w)}{(z-w)^2}\nonumber\\
&\qquad-\frac{n(n-1)\kappa(\kappa \tensor{\delta}{^c_b} \partial\tensor{U}{_{(1)}^a_d}(w)+\tensor{\delta}{^c_d} \partial\tensor{U}{_{(1)}^a_b}(w))}{(z-w)^2}\nonumber\\
&\qquad-\frac{(n-1)((\tensor{U}{_{(1)}^a_b}\tensor{U}{_{(1)}^c_d})(w)+\kappa (\tensor{U}{_{(1)}^a_d}\tensor{U}{_{(1)}^c_b})(w))}{(z-w)^2}\nonumber\\
&\qquad+\frac{\tensor{\delta}{^c_b} \tensor{U}{_{(3)}^a_d}(w)-\tensor{\delta}{^a_d} \tensor{U}{_{(3)}^c_b}(w)}{z-w}\nonumber\\
&\qquad-\frac{\kappa \tensor{\delta}{^c_b} \partial\tensor{U}{_{(2)}^a_d}(w)+\tensor{\delta}{^a_b} \partial\tensor{U}{_{(2)}^c_d}(w)}{z-w}\nonumber\\
&\qquad-\frac{n(n-1)\kappa(\kappa \tensor{\delta}{^c_b} \partial^2\tensor{U}{_{(1)}^a_d}(w)+\tensor{\delta}{^c_d} \partial^2\tensor{U}{_{(1)}^a_b}(w))}{2(z-w)}\nonumber\\
&\qquad+\frac{(\tensor{U}{_{(1)}^c_b}\tensor{U}{_{(2)}^a_d})(w)-\tensor{U}{_{(1)}^a_d}\tensor{U}{_{(2)}^c_b})(w)}{z-w}\nonumber\\
&\qquad-\frac{(n-1)(\kappa\partial\tensor{U}{_{(1)}^a_d}\tensor{U}{_{(1)}^c_b})(w)+\partial\tensor{U}{_{(1)}^a_b}\tensor{U}{_{(1)}^c_d})(w)}{z-w}\ .\label{eq:U2U2 OPE}
\end{align} \label{eq:low-lying OPEs}
\end{subequations}
For illustration, we have also written down the next OPEs $\tensor{U}{_{(1)}^a_b}(z)\tensor{U}{_{(4)}^c_d}(w)$ and $\tensor{U}{_{(2)}^a_b}(z)\tensor{U}{_{(3)}^c_d}(w)$ in Appendix~\ref{app:OPEs}.
Besides noticing that the algebra indeed seems to close on the higher spin fields $\tensor{U}{_{(j)}^a_b}(z)$, we observe many more special features:
\begin{enumerate}
\item First and foremost, it is highly surprising that the dependence on the rank $m$ of the matrix is completely absent in the OPE coefficients. This only seems to happen in the Miura basis. This is in contrast with the first OPEs in the primary basis worked out in \cite{Creutzig:2018pts}, where the dependence on $m$ is non-trivial. It is this property which allows us to proceed much further in our analysis than one might initially hope for. The flavor indices are never summed over in the OPE, only permuted. This is even more clear if we define $\tensor{U}{_{(0)}^a_b} = \tensor{\delta}{^a_b} \, \mathds{1}$ and use it to eliminate the Kronecker delta factors from the OPE. In other words, the only place where the rank $m$ enters the OPEs is the range of indices, but since we never sum of them, formally the algebra looks identical for any value of $m$.
\item We observe that the non-linearity is the Miura basis is at most quadratic, exactly as in the case of the Miura transformation in $\mathcal{W}_{1+\infty}$. Furthermore all the derivatives are determined completely by the coefficients of non-derivative terms and can be resummed if we consider bi-local expansions instead of OPEs. This is just as in \cite{Lukyanov:1988aa,Prochazka:2014gqa}.
\item A third special property of the Miura basis is that the coefficients are all polynomial in $\kappa$ and $n$. We will see below that the structure constants have also many symmetry properties.
\item Unlike the case of $\mathcal{W}_{1+\infty}$ where the fields $U_{(1)}$, $U_{(2)}$ and $U_{(3)}$ generate the full algebra, in the matrix-extended case already $U_{(1)}$ and $U_{(2)}$ suffice to generate the whole algebra.
\end{enumerate}
Since nothing in the operator product expansion depends on the rank $m$ of matrices or the structure of matrix multiplication (and this is true even for Jacobi identities), one expects that there should exist various generalizations of this construction. For super Lie algebras this is worked out explicitly in \cite{Rapcak:toappear}.

\subsection{Bootstrap}
We have computed the OPEs of $\tensor{U}{_{(j)}^a_b}(z)\tensor{U}{_{(j)}^c_d}(w)$ predicted by the Miura transformation for $j+k\le 7$. To explore the algebra to higher orders, it is useful to just \emph{impose} everything we know about the algebra and see that there is a unique solution. We refer to this technique as bootstrap and it was successfully applied to many $\mathcal{W}$-algebras \cite{Gaberdiel:2012ku, Candu:2012tr, Candu:2012ne, Beccaria:2013wqa, Beccaria:2014jra, Prochazka:2014gqa}. Essentially, we impose the Jacobi identities (i.e.~associativity) on the OPE, using the initial data provided by the Miura transformation. Using this technique, we have shown that there is a unique solution to the constraints coming from the Jacobi identities and have computed $\tensor{U}{_{(j)}^a_b}(z)\tensor{U}{_{(j)}^c_d}(w)$ for $j+k \le 16$ using Thieleman's package \cite{Thielemans:1991uw}. We will in the following sections discuss what we learn from the direct calculation.

\subsection{Structure constants}

Let us present the structure constants as found by the bootstrap calculation. We first concentrate on the non-derivative terms, which we write as
\begin{multline}
\tensor{U}{_{(j)}^a_b}(z)\tensor{U}{_{(k)}^c_d}(w) \sim \sum_{\ell+m < j+k} C_{j,k}^{\ell, m}(\kappa,n) \frac{\big(\tensor{U}{_{(\ell)}^a_b}\tensor{U}{_{(m)}^c_d}\big)(w)}{(z-w)^{j+k-l-m}} + \\
+ \sum_{\ell+m < j+k} \tilde{C}_{j,k}^{\ell, m}(\kappa,n) \frac{\big(\tensor{U}{_{(\ell)}^a_d}\tensor{U}{_{(m)}^c_b}\big)(w)}{(z-w)^{j+k-l-m}} + \text{derivatives}\ .\label{eq:OPE without derivatives}
\end{multline}
Our conjecture for the structure constants takes the form
\begin{align}
C_{j,k}^{0,0}(\kappa,n) & = \frac{(-1)^{k+1} \kappa^{j+k-2} (j+k-2)! n! (n-1)!}{(j-1)!(k-1)!(n-j)!(n-k)!} \nonumber\\
&\qquad\qquad\times\pFq{4}{3}{1-j, 1-k, 1-\kappa^{-1}, 1+\kappa^{-1}}{2, 2-j-k, 1-n}{1}\ ,\\
\tilde{C}_{j,k}^{0,0}(\kappa,n) & = \frac{(-1)^{k+1} \kappa^{j+k-1} (j+k-2)! n! (n-1)!}{(j-1)!(k-1)!(n-j)!(n-k)!} \nonumber\\
&\qquad\qquad\times\pFq{4}{3}{1-j, 1-k, 1-\kappa^{-1}, 1+\kappa^{-1}}{1, 2-j-k, 1-n}{1}\ ,
\end{align}
and for $\ell \ge 1, m=0$
\begin{align}
C_{j,k}^{\ell, 0}(\kappa,n) &= \frac{(-1)^{k+1} \kappa^{j+k-\ell-2} (n-\ell)!}{(n-j)!(n-k)!} \sum_{r=0}^{\min(j-2,k-1,j+k-\ell-2)} \!\!\!\!\!(j+k-\ell-r-2)!   \nonumber\\
&\qquad\times \prod_{s=1}^r \left( 1 - \frac{1}{\kappa^2 s^2} \right)\Bigg[ \frac{(n-r-1)!}{(r+1)(j-\ell-r-1)!(k-r-1)!} \nonumber\\
&\qquad+ \sum_{a=0}^{\ell-1} \sum_{b=0}^{\lfloor \frac{a-1}{2} \rfloor} \frac{(n-r-2-b)! (-1)^{a+1} {a-b-1 \choose b} {\ell-a+b \choose b+1} (r+2)_{b}}{(j-\ell-r+a-1-b)!(k-r-a-1+b)!} \Bigg]\ ,
\end{align}
and
\begin{align}
\tilde{C}_{j,k}^{\ell, 0}(\kappa,n) &= \delta_{j+k,\ell+1}+\frac{(-1)^{k+1} \kappa^{j+k-\ell-1} (n-\ell)!}{(n-j)!(n-k)!} \sum_{r=0}^{\min(j-1,k-1,j+k-\ell-2)}\!\!\!\!\! (j+k-\ell-r-2)!\nonumber\\
&\qquad\times \prod_{s=1}^r \left( 1 - \frac{1}{\kappa^2 s^2} \right)  \Bigg[ \frac{(n-r-1)!}{(j-\ell-r-1)!(k-r-1)!} \nonumber\\
&\qquad+ \sum_{a=0}^{\ell-1} \sum_{b=0}^{\lfloor \frac{a}{2} \rfloor} \frac{(-1)^a (n-r-b-1)! (r)_{b+1}{a-b \choose b}{\ell-a+b-1 \choose b}}{(j-\ell-r+a-b)!(k-r-a+b-1)!} \nonumber\\
&\qquad- \sum_{a=0}^{\ell-1} \sum_{b=0}^{\lfloor \frac{a-1}{2} \rfloor} \frac{(-1)^{a} (n-r-b-2)! (r+1)_{b+1} {a-b-1 \choose b} {\ell-a+b \choose b+1}}{(j-\ell-r+a-b-1)!(k-r-a+b -1)!} \Bigg]\ .
\end{align}
The structure constants for $\ell=0$ and $m \geq 1$ can be obtained from these using the symmetry
\begin{equation}
C_{j,k}^{\ell,m}(\kappa,n) = (-1)^{j+k-l-m} C_{k,j}^{m, \ell}(\kappa,n)
\end{equation}
and
\begin{equation}
\tilde{C}_{j,k}^{\ell, m}(\kappa,n) = (-1)^{j+k-l-m} \tilde{C}_{k,j}^{m, \ell}(\kappa,n)
\end{equation}
which is an obvious consequence of the commutativity of the OPE. To obtain structure constants with both $\ell$ and $m$ non-zero, we observe that the structure constants satisfy the translation symmetry
\begin{equation}
C_{j,k}^{\ell, m}(\kappa,n) = C_{j-1,k-1}^{\ell-1,m-1}(\kappa,n-1), \qquad j,\, k,\, \ell,\, m > 0\ ,
\end{equation}
and
\begin{equation}
\tilde{C}_{j,k}^{\ell, m}(\kappa,n) = \tilde{C}_{j-1,k-1}^{\ell-1,m-1}(\kappa,n-1), \qquad j,\, k,\, \ell,\, m > 0\ .
\end{equation}
The structure constants with $\ell  = m=0$ were derived in the same way as the corresponding structure constant in the case of $\mathcal{W}_{1+\infty}$ by solving a recurrence relation coming from the fusion coproduct \cite{Prochazka:2014gqa}. The remaining ones fit all the data obtained by bootstrap up to spins $j+k \leq 16$. Note that since nothing depends on the rank $m$ of the matrix part of the algebra, we could have used directly the results of \cite{Prochazka:2014gqa} in the $m=1$ case if only there was a way to distinguish the two index structures corresponding to $C^{\ell,m}_{j,k}(\kappa,n)$ and $\tilde{C}^{\ell,m}_{j,k}(\kappa,n)$. This is the only new element in the matrix-extended $\mathcal{W}_{1+\infty}$ algebra compared to standard $\mathcal{W}_{1+\infty}$. In $\mathcal{W}_{1+\infty}$ where $m=1$, both index structures become the same and $C + \tilde{C}$ corresponds to structure constants of \cite{Prochazka:2014gqa}.

To describe the derivative terms, we can essentially repeat the discussion in the case of $\mathcal{W}_{1+\infty}$ \cite{Prochazka:2014gqa}. All the derivative terms can be resummed if instead of standard OPEs one considers bi-local expansions. For each fixed $j$ and $k$, the combinations
\begin{multline}
\tensor{U}{_{(jk)}^{ac}_{bd}}(z,w) \equiv \tensor{U}{_{(j)}^a_b}(z)\tensor{U}{_{(k)}^c_d}(w) + \sum_{\ell+m < j+k} D_{j,k}^{\ell ,m}(\kappa,n) \frac{\tensor{U}{_{(\ell)}^a_b}(z)\tensor{U}{_{(m)}^c_d}(w)}{(z-w)^{j+k-l-m}}  \\
+ \sum_{\ell+m < j+k} \tilde{D}_{j,k}^{\ell, m}(\kappa,n) \frac{\tensor{U}{_{(\ell)}^a_d}(z)\tensor{U}{_{(m)}^c_b}(w)}{(z-w)^{j+k-l-m}}
\end{multline}
(valid for any finitely separated $z \neq w$, not just for the singular terms as $z \to w$) are regular as $z \to w$. The transformation between products of two $\tensor{U}{_{(j)}^a_b}$ fields and the bi-local fields $\tensor{U}{_{(jk)}^{ac}_{bd}}$ is triangular in the bi-index $(jk)$. So it can be inverted to give
\begin{multline}
\tensor{U}{_{(j)}^a_b}(z)\tensor{U}{_{(k)}^c_d}(w) = \sum_{\ell+m < j+k} C_{j,k}^{\ell, m}(\kappa,n) \frac{\tensor{U}{_{(\ell m)}^{ac}_{bd}}(z,w)}{(z-w)^{j+k-l-m}} + \\
+ \sum_{\ell+m < j+k} \tilde{C}_{j,k}^{\ell, m}(\kappa,n) \frac{\tensor{U}{_{(\ell m)}^{ac}_{db}}(z,w)}{(z-w)^{j+k-l-m}}\ ,
\end{multline}
which now holds as a bi-local expansion at separated points. Comparing this to \eqref{eq:OPE without derivatives} yields $D^{\ell,m}_{j,k}(\kappa,n)$ and $\tilde{D}^{\ell,m}_{j,k}(\kappa,n)$ and Taylor expanding this formula gives the usual operator product expansion, including all the derivative terms \cite{Prochazka:2014gqa}.

\subsection{The Virasoro tensor}
The algebra possesses a unique (total) Virasoro tensor with the following properties:
\begin{enumerate}
\item $\tensor{U}{_{(k)}^a_b}(z)$ has conformal weight $k$ w.r.t.~$T(z)$
\item All the spin $1$ currents $\tensor{U}{_{(1)}^a_b}(z)$ are primary of spin $1$ with respect to $T(z)$.
\end{enumerate}
It is given by
\begin{multline}
T(z)=\frac{1}{2(\kappa+m)} (\tensor{U}{_{(1)}^a_b}\tensor{U}{_{(1)}^b_a})(z)\\
+\frac{\kappa(n-1)}{2(\kappa+m)} \partial \tensor{U}{_{(1)}^a_a}(z)-\frac{1}{\kappa+m}\tensor{U}{_{(2)}^a_a}(z)\ .
\end{multline}
This can be seen as a generalized Sugawara construction. Indeed, for $n=1$, the latter two terms are not present and we recover the standard Sugawara construction. The central charge is as stated in \eqref{eq:central charge}.

\subsection{Coset description}\label{subsec:coset}
It is known \cite{Prochazka:2017qum,Eberhardt:2018plx, Creutzig:2018pts} that the (truncations of) matrix-extended $\mathcal{W}_{1+\infty}$ algebras can be realized as cosets of the form
\begin{equation}
\frac{\mathfrak{gl}(M+N)_k}{\mathfrak{gl}(N)_k}.
\end{equation}
Let us match these parameters with $(\kappa,n,m)$. First of all, our matrix-extended $\mathcal{W}_{1+\infty}$ has $\mathfrak{gl}(m)_{n\kappa}$ Kac-Moody symmetry while the coset has $\mathfrak{gl}(M)_k$ symmetry, i.e. we have
\begin{equation}
n\kappa = k, \quad\quad m=M.
\end{equation}
The central charge of the coset is
\begin{equation}
\frac{k((M+N)^2-1)}{k+M+N} - \frac{k(N^2-1)}{k+N}
\end{equation}
while our algebra has central charge \eqref{eq:central charge}. Comparing these we find two solutions,
\begin{subequations}
\begin{align}
m &= M, & n &= \frac{k}{k+N}, & \kappa &= k+N\ , \\
m &= M, & n &= -\frac{k}{k+M+N}, & \kappa& = -k-M-N
\end{align}
\end{subequations}
which are related by the duality \eqref{eq:duality symmetry}. Taking $M=m$ fixed and letting $N$ and $k$ run over positive integers we find a three one-parametric families of codimension $1$ truncations of matrix-extended $\mathcal{W}_{1+\infty}$,
\begin{align}
n \kappa - j &= 0\ , & n\kappa - \kappa + j &=0\ , & n\kappa + \kappa + m + j&=0\ ,  & j=1,\, 2,\, \ldots
\end{align}
Later we will discuss more general truncations and those we see here will be denoted by $jY, X+jZ$ and $X\tran+jZ$.

\subsection{Spectral flow}
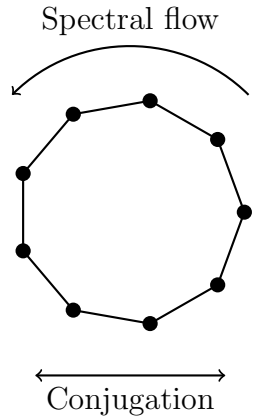
\begin{figure} 
\begin{center}
\begin{tikzpicture}
\fill (0:1.5) circle (.1);
\fill (40:1.5) circle (.1);
\fill (80:1.5) circle (.1);
\fill (120:1.5) circle (.1);
\fill (160:1.5) circle (.1);
\fill (200:1.5) circle (.1);
\fill (240:1.5) circle (.1);
\fill (280:1.5) circle (.1);
\fill (320:1.5) circle (.1);
\draw[thick] (0:1.5) -- (40:1.5) -- (80:1.5) -- (120:1.5) -- (160:1.5) -- (200:1.5) -- (240:1.5) -- (280:1.5) -- (320:1.5) -- (0:1.5);
\draw[thick,->] (45:2.2) arc (45:135:2.2) node[midway, above] {Spectral flow};
\draw[thick,<->] (240:2.5) to node[below] {Conjugation} (300:2.5);
\end{tikzpicture}
\end{center}
\caption{The Dynkin diagram of affine $\mathfrak{sl}(m)$ (in the example $m=9$) and the action of outer automorphisms.}\label{fig:affine glm dynkin diagram}
\end{figure}

Let us discuss a further discrete symmetry of the algebra. This symmetry is already present in the case of regular $\mathcal{W}_{1+\infty}[\lambda]$, but becomes much more interesting in the present context. The affine Kac-Moody algebra $\mathfrak{gl}(m)_k$ has the group $\mathrm{D}_m$ as outer automorphism group acting by spectral flow and conjugation (the latter is discussed in the next subsection). This is reflected in the symmetry of the affine Dynkin diagram, as displayed in Figure~\ref{fig:affine glm dynkin diagram}.

As we will discuss now, these symmetries extend to the whole of the matrix-extended $\mathcal{W}_{1+\infty}$ algebra.

Let us first recall the spectral flow of $\mathfrak{gl}(m)$. Geometrically, spectral flow is induced from conjugation of the Kac-Moody generators with loops in the loop group \cite{Pressley:1988qk, Fuchs:1992dda}. Let us consider the following loop in the loop group for $t \in [0,1]$
\be 
\gamma(t)=\mathrm{exp}(2\pi i t \boldsymbol{\omega})\in \mathrm{GL}(m)\ ,
\ee
where $\boldsymbol{\omega}=\mathrm{diag}(\omega_1,\dots,\omega_n)\in \mathfrak{gl}(m)$ is the spectral flow vector. In order for the loop to close, we need $\boldsymbol{\omega}\in \mathds{Z}^m$, which we can identify with the weight lattice of $\mathfrak{gl}(m)$. Such a loop acts as follows on the currents
\be 
\sigma^{\boldsymbol{\omega}}(\tensor{J}{^a_b})(z)=z^{-\omega_a+\omega_b} \tensor{J}{^a_b}(z)-\frac{\kappa\omega_a \tensor{\delta}{^a_b}}{z}\ .
\ee
By construction, this is an automorphism of the algebra (where we use the same normalizations as in the Miura transformation, see \eqref{eq:glm normalization convention}). 
If $\boldsymbol{\omega}$ is in the root lattice, the automorphism becomes inner, since we are conjugating with a contractible loop. Thus, there is a $\mathds{Z}_m$ spectral flow symmetry, which is outer. We can take it to be generated by $\boldsymbol{\omega}_0=(1,0,0,\dots,0)$. The outer automorphism acts also non-trivially on representations by performing a cyclic permutation of the affine Dynkin labels. Note however that this yields in general infinite-dimensional ground-state representations, since unless $k \in \mathds{Z}_{\ge 0}$, the resulting Dynkin labels associated to the finite-dimensional part of the algebra will not be integer.

To extend spectral flow symmetry to the matrix-extended $\mathcal{W}_{1+\infty}$ algebra, we simply apply spectral flow to each of the constituents in \eqref{eq:matrix Miura operator}. This leads to a very simple transformation law for the Miura operator:\footnote{In principle, one could imagine to choose $\boldsymbol{\omega}$ independently for each of the factors in the Miura transformation, but this does not preserve the algebra defined by the Miura transformation.}
\be 
\sigma^{\boldsymbol{\omega}}\left(\tensor{\mathcal{L}}{^a_b}\right)(z)
= z^{-\omega_a}\tensor{\mathcal{L}}{^a_b}(z) z^{\omega_b}\ .
\ee
Correspondingly, one can read off the transformation behavior of the higher spin fields
\be 
\sigma^{\boldsymbol{\omega}}(\tensor{U}{_{(j)}^a_b}(z))=\sum_{k=0}^j (-\kappa)^k {\omega_b \choose k} (n-j+1)_k \frac{z^{-\omega_a+\omega_b}\tensor{U}{_{(j-k)}^a_b}(z)}{z^k}\ ,
\ee
where $(a)_k$ is the Pochhammer symbol. Thus, the spectral flow symmetry readily extends to the matrix-extended $\mathcal{W}_{1+\infty}$ algebra. 
\subsection{Conjugation}
Finally, there is one last outer automorphism generator $\tau$. It again extends from $\mathfrak{gl}(m)_{n\kappa}$, where it is given by conjugation, sending the matrix to minus its transpose:
\be 
\tau(\tensor{U}{_{(1)}^a_b})(z)=-\tensor{U}{_{(1)}^b_a}(z)\ .
\ee
The extension to the matrix-extended $\mathcal{W}_{1+\infty}$ algebra takes the form
\be 
\tau(\tensor{U}{_{(j)}^a_b})(z)=(-1)^j \sum_{k=1}^j \binom{n-k}{j-k}\kappa^{j-k}\, \partial^{j-k} \tensor{U}{_{(k)}^b_a}(z)\ .\label{eq:conjugation action}
\ee
This can also be directly derived from the Miura transformation. Defining the action
\be
\tau(\tensor{J}{^{(i)}^a_b})=-\tensor{J}{^{(n+1-i)}^b_a}
\ee
on the generating fields, one can directly check that this automorphism induces the automorphism \eqref{eq:conjugation action} on the higher spin fields. 
Notice in particular the sign $(-1)^j$ in \eqref{eq:conjugation action}, which roughly implies that odd-spin fields get a minus sign under the automorphism.

Combining all the outer automorphisms we discussed (duality, spectral flow and conjugation), one can check directly that the full outer automorphism symmetry of the algebra is $\mathds{Z}_2 \times \mathrm{D}_m$, where $\mathrm{D}_m$ is the dihedral group with $2m$ elements. For $m=2$, conjugation is in fact inner, so the automorphism group is $\mathds{Z}_2\times \mathds{Z}_2$. Finally, for $m=1$, we have triality symmetry, so the automorphism group is $\mathrm{S}_3$.

\subsection{The block decomposition} \label{subsec:block decomposition}
In this section, we discuss the existence of a conformal embedding \footnote{We could just embed commuting diagonal elements of our matrices to find $m$ commuting copies of $\mathcal{W}_{1+\infty}$ algebra, but in order to make comparison with the gluing construction, we need a conformal embedding of these.}
\be
\mathcal{W}_{1+\infty}(\kappa,n;m_1) \times \mathcal{W}_{1+\infty}\left(\kappa+m_1,\tfrac{n\kappa}{\kappa+m_1}; m_2\right)
\subset \mathcal{W}_{1+\infty}(\kappa,n; m_1+m_2) \ ,\label{eq:embedding}
\ee
where $m=m_1+m_2$ and we wrote $\mathcal{W}_{1+\infty}(\kappa,n;m)$ for the matrix-extended $\mathcal{W}_{1+\infty}$ algebra with the respective parameters. We can define this embedding explicitly as follows. In the following, indices $a$, $b$, $\dots$ are indices of $\mathfrak{gl}(m_1)$ and run from $1$ to $m_1$, while $i$, $j$, $\dots$ are indices of $\mathfrak{gl}(m_2)$ and run from $m_1+1$ to $m_1+m_2=m$. We denote the corresponding fields by $\tensor{U}{^{[1]}_{(j)}^a_b}$ and $\tensor{U}{^{[2]}_{(k)}^i_j}$. We have
\begin{subequations}
\begin{align}
\tensor{U}{^{[1]}_{(j)}^a_b}&=\tensor{U}{_{(j)}^a_b}\ , \\
\tensor{U}{^{[2]}_{(1)}^i_j}&=\tensor{U}{_{(1)}^i_j}-\frac{\tensor{U}{_{(1)}^a_a}\tensor{\delta}{^i_j}}{\kappa+m_1}\ , \\
\tensor{U}{^{[2]}_{(2)}^i_j}&=\tensor{U}{_{(2)}^i_j}-\frac{\tensor{U}{_{(2)}^a_a}\tensor{\delta}{^i_j}+\left(\frac{\kappa}{2}+m_1\right)\partial\tensor{U}{_{(1)}^a_a}\tensor{\delta}{^i_j}}{\kappa+m_1}+m_1 \partial\tensor{U}{_{(1)}^i_j} -(\tensor{U}{_{(1)}^i_a}\tensor{U}{_{(1)}^a_j})\nonumber\\
&\qquad-\frac{(\tensor{U}{_{(1)}^a_a} \tensor{U}{_{(1)}^i_j})}{\kappa+m_1}+\frac{(\tensor{U}{_{(1)}^a_b} \tensor{U}{_{(1)}^b_a})\tensor{\delta}{^i_j}}{2(\kappa+m_1)}+\frac{(\tensor{U}{_{(1)}^a_a} \tensor{U}{_{(1)}^b_b})\tensor{\delta}{^i_j}}{2(\kappa+m_1)^2}\ , \\
\tensor{U}{^{[2]}_{(3)}^i_j}&= \tensor{U}{_{(3)}^i_j}-\frac{\tensor{U}{_{(3)}^a_a}\tensor{\delta}{^i_j}+m_1\partial\tensor{U}{_{(2)}^a_a}\tensor{\delta}{^i_j}+\frac{1}{6}(3m_1^2-3+4m_1\kappa+2\kappa^2)\partial^2\tensor{U}{_{(1)}^a_a}\tensor{\delta}{^i_j}}{\kappa+m_1}\nonumber\\
&\qquad+m_1 \partial\tensor{U}{_{(2)}^i_j}+\frac{m_1(\kappa+m_1-1)(\kappa+m_1+1)\partial^2\tensor{U}{_{(1)}^i_j}}{2(\kappa+m_1)}-(\tensor{U}{_{(1)}^i_a}\tensor{U}{_{(2)}^a_j}) \nonumber\\
&\qquad-(\tensor{U}{_{(1)}^a_j}\tensor{U}{_{(2)}^i_a}) -\frac{(\tensor{U}{_{(1)}^a_a}\tensor{U}{_{(2)}^i_j})+(\tensor{U}{_{(1)}^i_j}\tensor{U}{_{(2)}^a_a})-(\tensor{U}{_{(1)}^a_b}\tensor{U}{_{(2)}^b_a})\tensor{\delta}{^i_j} }{\kappa+m_1}\nonumber\\
&\qquad +\frac{(\tensor{U}{_{(1)}^a_a}\tensor{U}{_{(2)}^b_b})\tensor{\delta}{^i_j}}{(\kappa+m_1)^2} -(\kappa+2m_1) (\tensor{U}{_{(1)}^i_a}\partial\tensor{U}{_{(1)}^a_j})\nonumber\\
&\qquad-\frac{(\kappa+2m_1) (\tensor{U}{_{(1)}^a_a}\partial\tensor{U}{_{(1)}^i_j})}{\kappa+m_1}-\frac{(m_1^2+1+\kappa m_1) (\tensor{U}{_{(1)}^a_j}\partial\tensor{U}{_{(1)}^i_a})}{\kappa+m_1}\nonumber\\
&\qquad-\frac{(\kappa+2m_1)(\tensor{U}{_{(1)}^i_j}\partial\tensor{U}{_{(1)}^a_a})}{2(\kappa+m_1)}+\frac{(3\kappa+5m_1)(\tensor{U}{_{(1)}^a_b}\partial\tensor{U}{_{(1)}^b_a})\tensor{\delta}{^i_j}}{3(\kappa+m_1)}\nonumber\\
&\qquad+\frac{(5\kappa+8m_1)(\tensor{U}{_{(1)}^a_a}\partial\tensor{U}{_{(1)}^b_b})\tensor{\delta}{^i_j}}{6(\kappa+m_1)^2}+(\tensor{U}{_{(1)}^i_a}\tensor{U}{_{(1)}^a_b}\tensor{U}{_{(1)}^b_j})\nonumber\\
&\qquad+\frac{(\tensor{U}{_{(1)}^i_a}\tensor{U}{_{(1)}^b_b}\tensor{U}{_{(1)}^a_j})+\frac{1}{2} (\tensor{U}{_{(1)}^a_b}\tensor{U}{_{(1)}^b_a}\tensor{U}{_{(1)}^i_j})-\frac{1}{3} (\tensor{U}{_{(1)}^a_b}\tensor{U}{_{(1)}^b_c}\tensor{U}{_{(1)}^c_a})\tensor{\delta}{^i_j}}{\kappa+m_1}\nonumber\\
&\qquad+\frac{(\tensor{U}{_{(1)}^a_a}\tensor{U}{_{(1)}^b_b}\tensor{U}{_{(1)}^i_j})-(\tensor{U}{_{(1)}^a_b}\tensor{U}{_{(1)}^b_a}\tensor{U}{_{(1)}^c_c})\tensor{\delta}{^i_j}}{2(\kappa+m_1)^2}\nonumber\\
&\qquad-\frac{(\tensor{U}{_{(1)}^a_a}\tensor{U}{_{(1)}^b_b}\tensor{U}{_{(1)}^c_c})\tensor{\delta}{^i_j}}{6(\kappa+m_1)^3}\ .
\end{align}
\end{subequations}
Since $\tensor{U}{^{[2]}_{(1)}^i_j}$, $\tensor{U}{^{[2]}_{(2)}^i_j}$ and $\tensor{U}{^{[2]}_{(3)}^i_j}$ generate the matrix-extended $\mathcal{W}_{1+\infty}$ algebra, all higher generators are also fixed.\footnote{In fact, for $m_2>1$ already $\tensor{U}{^{[2]}_{(1)}^i_j}$ and $\tensor{U}{^{[2]}_{(2)}^i_j}$ generate the complete algebra.}
One can check directly that $\tensor{U}{^{[1]}_{(j)}^a_b}$ and $\tensor{U}{^{[2]}_{(k)}^i_j}$ commute and the lowest $\tensor{U}{^{[2]}_{(k)}^i_j}$ satisfy the OPEs of the matrix-extended $\mathcal{W}_{1+\infty}$ with parameters given in \eqref{eq:embedding}.

One can iterate this procedure to find the embedding
\be 
\bigtimes_{\ell=1}^m \, \mathcal{W}_{1+\infty}\left(\kappa+\ell,\tfrac{n\kappa}{\kappa+\ell}\right)\subset\mathfrak{gl}(m) \ \mathcal{W}_{1+\infty}(\kappa,n) \ .\label{eq:iterated embedding}
\ee
Thus, there are $m$ commuting $\mathcal{W}_{1+\infty}$ subalgebras inside matrix-extended $\mathcal{W}_{1+\infty}$. The corresponding $\lambda$-parameters read
\be 
\lambda^\ell_1=\frac{n\kappa}{\kappa+\ell-1}\ ,\qquad \lambda_2^\ell=-\frac{n\kappa}{\kappa+\ell}\ , \qquad \lambda_3^\ell=n \kappa\ . \label{eq:lambda ell parameters}
\ee

\section{Truncations of algebra} \label{sec:representations}
Finally, we discuss truncations of the algebra.
We have already seen several such truncations. For $n \in \mathds{N}$, it is manifest from the Miura transformation that the field $\tensor{U}{_{(n+1)}^a_b}(z)$ is never produced in OPEs and hence generates an ideal inside the algebra. Similarly, it is well-known that in $\mathfrak{gl}(m)_k$ with $k \in \mathds{Z}$, the field
\be 
\underbrace{(\tensor{U}{_{(1)}^1_m}\cdots \tensor{U}{_{(1)}^1_m})}_{k+1}
\ee
generates an ideal inside the algebra. We have also seen a third family of truncations using the coset description in Section~\ref{subsec:coset}.
However, there are more general truncations which lead to an intriguing structure generalizing the Y-algebras to the matrix case.
\subsection{Highest weight states}
To discuss the representation theory below, it will be useful to understand higher-weight states of the algebra. To do so, it is simplest to think about the algebra in terms of its modes. The OPE of the fields like \eqref{eq:low-lying OPEs} can be translated into commutators of modes by computing
\begin{align}
[\tensor{U}{_{(j)}^a_{b,r}},\tensor{U}{_{(k)}^c_{d,s}}]=\oint_0 \mathrm{d}w \oint_w \mathrm{d}z\ z^{r+j-1}w^{s+k-1} \tensor{U}{_{(j)}^a_b}(z)\tensor{U}{_{(k)}^c_d}(w)\ .
\end{align}
A highest-weight representation of the algebra has a highest-weight state $\ket{\lambda}$ satisfying
\be 
\tensor{U}{_{(j)}^a_{b,r}}\ket{\lambda}=0\ . \label{eq:highest weight condition}
\ee
for all $j\in \mathds{N}$, all $a$ and $b$ and all $r>0$. We also demand that the highest weight state is a highest weight state with respect to the global $\mathfrak{gl}(m)$ subalgebra, which amounts to
\be 
\tensor{U}{_{(1)}^a_{b,0}}\ket{\lambda}=0
\ee
for $a<b$. This makes the highest weight state unique.

One can see from the OPEs, that it is sufficient to impose the conditions
\begin{subequations}
\begin{align} 
\tensor{U}{_{(1)}^i_{i+1,0}} \ket{\lambda}&=0\ , \qquad i \in \{1,\dots,m-1\}\ , \\
\tensor{U}{_{(1)}^m_{1,1}} \ket{\lambda}&=0\ , \\
\tensor{U}{_{(1)}^1_{1,1}} \ket{\lambda}&=0\ , \\
\tensor{U}{_{(2)}^m_{1,1}} \ket{\lambda}&=0\ , \\
\tensor{U}{_{(3)}^m_{1,1}} \ket{\lambda}&=0\ , 
\end{align}\label{eq:simple highest weight condition}
\end{subequations}
Thus, these generators can be thought of as the `simple roots' of the algebra.

The Verma module is then generated by acting with all negative modes, as well as with $\tensor{U}{_{(0)}^a_b}$ for $a>b$ on the highest weight state. The irreducible representation space is obtained by dividing the Verma module by all null-vectors. 

\subsection{Singular fields}
We now consider the algebra for low values of $m=2,3,4,5$ and we compute all truncations up to certain level. Technically, we do this by computing singular fields, i.e.~fields which are both descendants and highest-weight fields satisfying \eqref{eq:simple highest weight condition}. By systematically listing the states in the vacuum representation of the algebra, we can search for singular vectors satisfying these conditions. In this brute-force analysis, we are restricted to small $m$. For $m=2$, we have been able to work up to level 8, for $m=3$, $4$ up to level 6, and for $m=5$ up to level 5. Equivalently, one could work out the Kac determinant (Shapolov form) of the vacuum representation. However, this is computationally far more ineffective, so we chose to work with singular fields instead.

As expected, we find that singular fields appear only if certain relations beween the parameters $\kappa, n$ and $m$ are satisfied. For instance, at level 2, we find four different curves in the parameter space where singular fields appear. These are 
\be 
n-1=0\ , \qquad n \kappa+\kappa+m=0\ , \qquad n \kappa-1=0\ , \qquad n \kappa+1=0\ . 
\ee
The first and third curves were expected, and already discussed in the introductory part of this section. The presence of the second curve is required by the duality symmetry \eqref{eq:duality symmetry}. The last curve is already duality invariant on its own. One can think of these four truncations as corresponding to four basic Miura factors. Analogously to the situation in $\mathcal{W}_{1+\infty}$, applying the fusion to these basic Miura transforms allows us to realize any other truncation of the algebra. The first Miura factor is the central object studied here and represents matrix-valued $\mathcal{W}_{1+\infty}$ in terms of a Kac-Moody algebra. The second one is obtained from it by applying duality automorphism as already mentioned. The last two elementary Miura factors are studied in greater detail in \cite{Rapcak:toappear}.

We should also mention that the latter three curves in parameter space degenerate for $n=1$ to the singular fields of $\mathfrak{gl}(m)_\kappa$ Kac-Moody algebra (provided that $m \ge 3$). The complete data of our computation is discussed in the next subsection.
\subsection{The gluing construction}
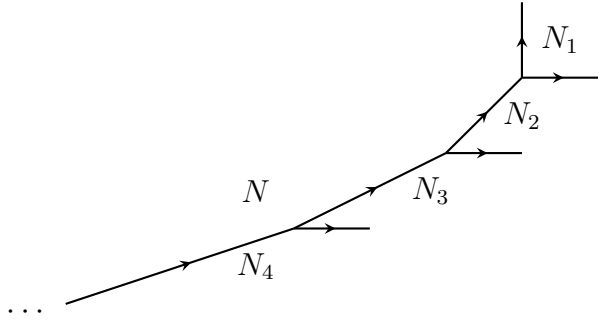
\begin{figure}
\begin{center}
\begin{tikzpicture}
\draw[thick,directed] (0,0) -- (0,1);
\draw[thick,directed] (-1,-1) -- (0,0);
\draw[thick,directed] (-3,-2) -- (-1,-1);
\draw[thick,directed] (-6,-3) -- (-3,-2);
\draw[thick, directed] (0,0) -- (1,0);
\draw[thick, directed] (-1,-1) -- (0,-1);
\draw[thick, directed] (-3,-2) -- (-2,-2);
\node at (-6.5,-3.1) {$\cdots$};
\node at (.5,.5) {$N_1$};
\node at (0,-.5) {$N_2$};
\node at (-1.2,-1.5) {$N_3$};
\node at (-3.5,-2.5) {$N_4$};
\node at (-3.5,-1.5) {$N$};
\end{tikzpicture}
\end{center}
\caption{Truncations of the matrix-extended $\mathcal{W}_{1+\infty}$ algebra via gluing.} \label{fig:matrix-extended Winfty gluing}
\end{figure}

In order to interpret the results from our computation, it is useful to employ the gluing construction, which we have reviewed in Subsection~\ref{subsec:truncations and gluing}. 
The $m$ commuting $\mathcal{W}_{1+\infty}$ algebras inside matrix-extended $\mathcal{W}_{1+\infty}$, which were discussed in Subsection~\ref{subsec:block decomposition} are exactly those $\mathcal{W}_{1+\infty}$ algebras whose Y-algebras can be glued together according to the gluing construction \cite{Prochazka:2017qum}. Thus, we expect that truncations of the matrix-extended version of $\mathcal{W}_{1+\infty}$ are captured by gluings as in Figure~\ref{fig:matrix-extended Winfty gluing}. We will see that these gluings account in fact for \emph{all} truncations of the matrix-extended algebra that we found by explicit study of singular vectors.
In order for the gluing to be possible, we need to satisfy all the gluing conditions \eqref{eq:gluing conditions}. Moreover, we want to impose the additional condition
\be 
N_{\ell+1}-N_\ell=N_{\ell+2}-N_{\ell+1}\ ,\qquad \ell \in \{1,\dots,m-1\}\ ,
\ee
which ensures that the basic gluing fields have spin 1. Hence, the numbers on the right of the Figure~\ref{fig:matrix-extended Winfty gluing} change linearly. Then the truncation condition \eqref{eq:truncation condition} is equivalent at every vertex:
\be 
\frac{N_{\ell+1}}{\lambda_1^\ell}+\frac{N_\ell}{\lambda_2^\ell}+\frac{N}{\lambda_3^\ell}=\frac{(\kappa+\ell-1) N_{\ell+1}-(\kappa+\ell) N_\ell+N}{n\kappa}\overset{!}{=}1\ . \label{eq:null locus condition}
\ee
From the direct search for singular vectors, it turns out that the truncations predicted from the gluing construction account for \emph{all} singular vectors. From now we discuss the truncations of the algebra for rank $2$, i.e. $m=2$. All the singular vectors for $m=2$ up to level 8 are displayed in Table~\ref{tab:truncations m=2}.\footnote{We have omitted level 1, since it behaves a bit singular in our parametrization. We have a null-vector at level $1$, if $n\kappa=0$. But since the denominator in \eqref{eq:null locus condition} vanishes at this value, we cannot directly translate this condition into a gluing picture. Morally, it should however correspond to the diagram, where all entries are zero. 
The resulting truncation is the Heisenberg algebra.}

\afterpage{
\renewcommand{\arraystretch}{4}
\begin{longtable}[c]{c|cccc}
Level 2 & \gluingtwo{2}{1}{0}{0}{0}{$n \kappa-1$}{2} & \gluingtwo{2}{0}{2}{1}{0}{$n -1$}{1} & \gluingtwo{2}{0}{0}{1}{2}{$n \kappa+\kappa+2$}{1} & \gluingtwo{2}{0}{1}{1}{1}{$n \kappa+1$}{0} \\
\hline
Level 3 & \gluingtwo{3}{2}{0}{0}{0}{$n \kappa-2$}{3} & \gluingtwo{3}{0}{4}{2}{0}{$n -2$}{1} & \gluingtwo{3}{0}{0}{2}{4}{$n \kappa+2\kappa+4$}{1}  \\
\hline
\multirow{2}{*}{\vspace{-1cm} Level 4} & \gluingtwo{4}{3}{0}{0}{0}{$n \kappa-3$}{4} & \gluingtwo{4}{1}{2}{1}{0}{$n\kappa-\kappa-1$}{2} & \gluingtwo{4}{1}{0}{1}{2}{$n\kappa+\kappa+1$}{2} & \gluingtwo{4}{0}{6}{3}{0}{$n-3$}{1}  \\
\nopagebreak
 & \gluingtwo{4}{0}{0}{3}{6}{$n \kappa+3\kappa+6$}{1} & \gluingtwo{4}{0}{1}{2}{3}{$n\kappa+\kappa+3$}{0} & \gluingtwo{4}{0}{3}{2}{1}{$n\kappa-\kappa+1$}{0} \\
\hline
Level 5 & \gluingtwo{5}{4}{0}{0}{0}{$n \kappa-4$}{5}  & \gluingtwo{5}{0}{8}{4}{0}{$n-4$}{1}  & \gluingtwo{5}{0}{0}{4}{8}{$n\kappa+4\kappa+8$}{1} & \gluingtwo{5}{0}{2}{2}{2}{$n\kappa+2$}{1} \\
\hline
\multirow{3}{*}{\vspace{-2cm} Level 6} &\gluingtwo{6}{5}{0}{0}{0}{$n \kappa-5$}{6} & \gluingtwo{6}{2}{2}{1}{0}{$n \kappa-\kappa-2$}{3} & \gluingtwo{6}{2}{0}{1}{2}{$n\kappa+\kappa$}{3} & \gluingtwo{6}{1}{4}{2}{0}{$n \kappa-2\kappa-1$}{2}  \\
\nopagebreak
& \gluingtwo{6}{1}{0}{2}{4}{$n \kappa+2\kappa+3$}{2}& \gluingtwo{6}{0}{10}{5}{0}{$n-5$}{1} & \gluingtwo{6}{0}{0}{5}{10}{$n \kappa+5\kappa+10$}{1} & \gluingtwo{6}{0}{5}{3}{1}{$n \kappa-2\kappa+1$}{0} \\
\nopagebreak
& \gluingtwo{6}{0}{1}{3}{5}{$n \kappa+2\kappa+5$}{0} \\
\hline
Level 7 & \gluingtwo{7}{6}{0}{0}{0}{$n \kappa-6$}{7}  & \gluingtwo{7}{0}{12}{6}{0}{$n-6$}{1}  & \gluingtwo{7}{0}{0}{6}{12}{$n\kappa+6\kappa+12$}{1}  \\
\hline
\multirow{3}{*}{\vspace{-2cm} Level 8} &\gluingtwo{8}{7}{0}{0}{0}{$n \kappa-7$}{8} & \gluingtwo{8}{3}{2}{1}{0}{$n \kappa-\kappa-3$}{4} & \gluingtwo{8}{3}{0}{1}{2}{$n+\kappa-1$}{4} & \gluingtwo{8}{1}{6}{3}{0}{$n \kappa-3\kappa-1$}{2}  \\
\nopagebreak
& \gluingtwo{8}{1}{0}{3}{6}{$n \kappa+3\kappa+5$}{2}& \gluingtwo{8}{0}{14}{7}{0}{$n-7$}{1} & \gluingtwo{8}{0}{0}{7}{14}{$n \kappa+7\kappa+14$}{1} & \gluingtwo{8}{0}{4}{3}{2}{$n \kappa-\kappa+2$}{1} \\
\nopagebreak
& \gluingtwo{8}{0}{2}{3}{4}{$n \kappa+\kappa+4$}{1} & \gluingtwo{8}{0}{7}{4}{1}{$n \kappa-3\kappa+1$}{0} & \gluingtwo{8}{0}{1}{4}{7}{$n \kappa+3\kappa+7$}{0} & \gluingtwo{8}{0}{3}{3}{3}{$n \kappa+3$}{0} \vspace{-.5cm}\\
\caption{Truncations for $m=2$. We suppressed the angles and arrows in the gluing diagrams for simplicity.} \label{tab:truncations m=2}
\end{longtable}
\renewcommand{\arraystretch}{1}
}

We should note that in this picture, the duality symmetry \eqref{eq:duality symmetry} simply acts by vertically reflecting the diagram. The fact that truncations always appear in duality invariant pairs (or are duality invariant by themselves) is a good consistency check of our computations.

We have decorated some vertices with red dots. These signal that the respective vertex already possesses a null-vector at the given level. Indeed, the Y-algebra in the vertex has a null-vector at level \eqref{eq:null-vector level}, which remains null in the extended algebra. Experimentally, we see that this always happens if $N_1=0$ or $N_3=0$. We observe in these cases that there is also a very simple rule to determine the $\mathfrak{sl}(2)$ representation in which the null-vector resides: The spin is always one more than the left number in the diagram. We should also clarify that the null-vector appears for all spins up to some maximal spin, which is the spin we display in Table~\ref{tab:truncations m=2}.

There are however also cases without red dots, like the last diagram for the level 2 truncations in Table~\ref{tab:truncations m=2}. While for these values in parameter space, the vertices do possess null-vectors, they would appear at higher levels (and different spins). Thus, the singular vector originates from the gluing fields in these cases. For instance, focusing again on the last diagram for level 2, the vertices do contain null-vectors, but by \eqref{eq:null-vector level} these would only appear at level $4=(0+1)(1+1)(1+1)$.

Let us look in more detail at the level 2 singular vector. There is a unique field which is primary with respect to both the commuting $\mathcal{W}_{1+\infty}$ subalgebras with $\widehat{\mathfrak{u}}(1)$ charge 0.\footnote{Note that our usage of the two commuting $\mathcal{W}_{1+\infty}$ algebras breaks $\mathfrak{gl}(2) \to \mathfrak{u}(1) \oplus \mathfrak{u}(1)$. Since the overall $\mathfrak{u}(1)$ charge is still central, we measure here the $\mathfrak{u}(1) \subset \mathfrak{sl}(2)$ charge.} It represents the gluing primary which is with respect to both of these algebras a primary labeled by $(\Box, \overline{\Box})$ representation along the glued edge.
\begin{align}
U^\text{prim}_{(2)}&=-n (n\kappa-\kappa+1) \tensor{U}{_{(2)}^1_1}+n (\kappa +\kappa  n+1) \tensor{U}{_{(2)}^2_2}+\kappa ^2 (n-1) n\,
   \partial \tensor{U}{_{(1)}^1_1}\nonumber\\
&\qquad -\kappa n (n-1) (\kappa  n+\kappa+1) (\tensor{U}{_{(1)}^1_2}\tensor{U}{_{(1)}^2_1})+(n-1) (\kappa  n+1) (\tensor{U}{_{(1)}^1_1}\tensor{U}{_{(1)}^1_1})\nonumber\\
   &\qquad-(n-1) (\kappa n+\kappa +1) (\tensor{U}{_{(1)}^1_1}\tensor{U}{_{(1)}^2_2})\ .
\end{align}
It becomes the singular field upon specification $n\kappa-1=0$, $n-1=0$, $n \kappa+\kappa+2=0$ or $n\kappa+1=0$. This primary field does not sit in a definite $\mathfrak{sl}(2)$ representation, but upon specification of the parameters sits either in the spin 0, spin 1 or spin 2 representation, according to Table~\ref{tab:truncations m=2}.
\subsection{The stable range}
After having discussed the case $m=2$, let us discuss also higher values $m$. It is particularly useful to look at region of large $m$, since then some irregularities associated to small matrix size disappear and the properties of singular vectors stabilize.

We have checked that the same structure persists for higher values of $m$ and the corresponding data for $m=3$, $4$ and $5$ is shown in Appendix~\ref{app:null-vectors higher m}. One can see that the singular gluing fields appear here at different levels compared to their analogues for $m=2$. For example, let us look at the sequence of truncation curves
\be 
\gluingtwobare{5}{0}{3}{3}{3}\ , \qquad \gluingthreebare{5}{0}{3}{3}{3}{3} \ , \qquad \gluingfourbare{5}{0}{3}{3}{3}{3}{3} \ , \qquad \gluingfivebare{5}{0}{3}{3}{3}{3}{3} \ , \dots
\ee
The first null-vector appears at level 8 for $m=2$, but at level 6 for $m=3$ and for $m \ge 4$, it appears at level 4. We define the stable level to be the level of the null-vector for $m \to \infty$. Note however that the corresponding null-vector appears at this level for all but a finite number of values of $m$. For small values of $m$, the corresponding representation might not exist, which is why one observes the null-vector only higher up. Describing this unstable behavior is complicated and we focus in the following on the stable levels.

To formalize the discussion, it is useful to introduce names for the four truncation conditions appearing at level 2 (here displayed for $m=4$):
\be 
X=\gluingfourbare{2}{0}{4}{3}{2}{1}{0}\ , \qquad X\tran=\gluingfourbare{2}{0}{0}{1}{2}{3}{4}\ ,\qquad Y=\gluingfourbare{2}{1}{0}{0}{0}{0}{0}\ , \qquad Z=\gluingfourbare{2}{0}{1}{1}{1}{1}{1}\ .
\ee
Note that $X$ and $X\tran$ are exchanged under duality. They represent our basic Miura factor and its dual.
The space of all allowed truncations is then the positive cone over these four generators, i.e.~every truncation curve can be written as
\be 
k_X X + k_{X\tran} X\tran + k_Y Y + k_Z Z
\ee
with $k_X$, $k_{X\tran}$, $k_Y$ and $k_Z \in \mathds{Z}_{\ge 0}$. The corresponding truncation curve in the parameter space reads
\be 
n\kappa-k_X\kappa+k_{X\tran}(\kappa+m)-k_Y+k_Z=0\ .
\ee
Since we can shift all numbers in the gluing diagram equally (without changing the values of parameters of the algebra), we can identify $Y+Z \sim 0$ and thus always choose either $k_Y=0$ or $k_Z=0$. Similarly, we have $X+X\tran=m Z$, which allows us to choose either $k_X=0$ or $k_{X\tran}=0$.

One observes from the data that $k_Y$ and $k_Z$ always have the null-vector at the same stable level $k+1$. Thus, there seems to be a second `duality' symmetry\footnote{We put quotation marks around it, since this is not a real symmetry.} in the large rank limit which exchanges the truncation $Y$ with $Z$. This lets one make a guess for the level at which the null-vector appears.

If $k_Z=0$, we always have a read dot present in the diagram and thus the level of the null-vector is inherited from $\mathcal{W}_{1+\infty}$ \cite{Prochazka:2017qum}. This leads to the formula
\be 
\text{Level}=(k_X+1)(k_{X\tran}+1)(k_Y+1)\ ,
\ee
where it is understood that either $k_X=0$ or $k_{X\tran}=0$.

If on the other hand $k_Y=0$, we expect the same level by the additional `duality' symmetry, except that $k_Y$ is interchanged with $k_Z$. Thus, we conjecture the general formula for the level of the null-vector in the stable range to be
\be 
\text{Level}=(k_X+1)(k_{X\tran}+1)(k_Y+1)(k_Z+1)\ ,
\ee
where it is again understood that either $k_X=0$ or $k_{X\tran}=0$ and similarly that either $k_Y=0$ or $k_Y=0$. This conjecture is consistent with all the data we have computed. The additional `duality' symmetry and the elegant final formula seems to beg for a deeper interpretation.\footnote{At the level of Grassmannian subalgebra of matrix-valued $\mathcal{W}_{1+\infty}$ (which is obtained by taking the coset with respect to the $\mathfrak{gl}(m)_{n\kappa}$ subalgebra), it is a consequence of the exact duality which exchanges all the ranks with their negatives (which is actually what we see in these diagrams). Having a finite value of rank $m=2,3,\ldots$ however breaks this symmetry so we see its consequences only in the stable region where the value of the rank $m$ does not play a role.}

One can similarly make an educated guess for the $\mathfrak{gl}(m)$ representation which the null-state appears. From the data, the representation of the null-vector seems to be given by the Dynkin labels
\be 
[k_Y+1,0,\dots,0,k_Y+1]
\ee
in the stable range.

\section{Summary and outlook} \label{sec:discussion}

The Miura transformation in the case of $\mathcal{W}_{1+\infty}$ is a very powerful construction. Apart from providing a free field representation of the algebra, it also shows the existence of a quadratic basis of the generating fields which allows one to write the operator product expansions of $\mathcal{W}_{1+\infty}$ in a closed form. In the first part of this work we checked that all these special properties remain true even if we consider the matrix-extended version of $\mathcal{W}_{1+\infty}$. The free field representation is now replaced by an affine Lie algebra, but the operator product expansions remain quadratic and again we are able to write closed form formulas for all the OPEs. Surprisingly, there is almost no dependence of the OPEs on the rank or structure of the matrix part of the algebra.

In the second part we studied the structure of truncations of the vacuum module of the algebra which shows a more intricate structure than in the case of $\mathcal{W}_{1+\infty}$. The gluing construction explains for which codimension $1$ curves in the parameter space there is a singular vector in the vacuum representation, but to understand the level and spin of these one must make a more detailed analysis. The space of simple truncations turns out to be a cone over positive integers generated by four basic level $2$ truncations.

There are several possible further directions and applications that are worth studying.

\paragraph{Yangian, $R$-matrix, shuffle algebra}
The algebra $\mathcal{W}_{1+\infty}$ has a dual description as the affine Yangian of $\widehat{\mathfrak{gl}}(1)$ \cite{Tsymbaliuk:2014fvq}. One way to understand the map to Yangian variables is via the Maulik-Okounkov $R$-matrix \cite{Maulik:2012wi, Zhu:2015nha, Prochazka:2019dvu}. It would be very interesting to generalize this construction to the matrix-extended case and find the matrix-extended version of the Yangian generators, see also \cite{Costello:2017fbo}. The generators of $\mathcal{W}_{1+\infty}$ act on orbital degrees of freedom in Calogero models \cite{schiffmann2013cherednik}, so one expects there should be an action of matrix-extended $\mathcal{W}_{1+\infty}$ on spin-Calogero models. The Yangian variables are closely related to the shuffle algebra description and it would be nice to have a matrix generalization of the map \cite{Negut:2016dxr} between the quadratic basis generators and the shuffle algebra.

\paragraph{Grassmannian}
Although $\mathcal{W}_{1+\infty}$ can be used as a building block via the gluing procedure to construct a large class of vertex operator algebras, there are examples of vertex operator algebras which don't seem to be decomposable into these blocks. One such example is the Grassmannian coset $\mathrm{GL}(M+N)_k/\mathrm{GL}(M)_k \times \mathrm{GL}(N)_k$ which for $M,N>1$ is larger than $\mathcal{W}_{1+\infty}$ and in fact there is a three-parametric family of these algebras (i.e. we have one more parameter than in $\mathcal{W}_{1+\infty}$). This algebra is a subalgebra of the matrix-extended $\mathcal{W}_{1+\infty}$ and should have a duality symmetry $\mathrm{S}_3 \times \mathds{Z}_2$. It is the fundamental building block of all unitary flag coset algebras. All these properties make it an interesting object to study.

\section*{Acknowledgements}
We thank Miroslav Rap\v{c}\'{a}k for useful discussions as well as for sharing the draft \cite{Rapcak:toappear} prior to publication. we would also like to thank to Branislav Jur\v{c}o, Yasuaki Hikida, Wei Li, Andy Linshaw, Shashank Kanade, Luca Matiello, Cheng Peng and Ivo Sachs for useful discussions. We thank Alessandro Sfondrini for organizing the workshop ``A fresh look at AdS$_3$/CFT$_2$'' in Castasegna, where this work started. We would also like to thank the Erwin Schr\"odinger Institute in Vienna, where the first part of this work was completed, for hospitality.
LE is supported by the Swiss National Science Foundation, and by the NCCR SwissMAP which is also funded by the Swiss National Science Foundation. He also acknowledges support from the Della Pietra Family at IAS.
The research of TP was supported by the DFG Transregional Collaborative Research Centre TRR 33 and the DFG cluster of excellence Origin and Structure of the Universe.

\appendix
\section{OPEs and structure constants}\label{app:OPEs}
In the main text, we have displayed the OPEs of the algebra until $j+k=4$. Here, we display for further illustration of the properties discussed in the main text also the OPEs with $j+k=5$. They take the form
\begin{align}
\tensor{U}{_{(1)}^a_b}&(z)\tensor{U}{_{(4)}^c_d}(w) \nonumber\\
&\sim-\frac{(n-3) (n-2) (n-1) n  \kappa ^3(\kappa  \tensor{\delta}{^a_d} \tensor{\delta}{^c_b}+\tensor{\delta}{^a_b} \tensor{\delta}{^c_d})}{(z-w)^5}\nonumber\\
&\qquad+\frac{(n-3) (n-2) (n-1) \kappa ^2(\kappa  \tensor{\delta}{^a_d} \tensor{U}{_{(1)}^c_b}(w)+\tensor{\delta}{^a_b} \tensor{U}{_{(1)}^c_d}(w)) }{(z-w)^4}\nonumber\\
&\qquad-\frac{(n-3) (n-2) \kappa(\kappa  \tensor{\delta}{^a_d} \tensor{U}{_{(2)}^c_b}(w)+\tensor{\delta}{^a_b} \tensor{U}{_{(2)}^c_d}(w))  }{(z-w)^3}\nonumber\\
&\qquad+\frac{(n-3) (\kappa  \tensor{\delta}{^a_d} \tensor{U}{_{(3)}^c_b}(w)+\tensor{\delta}{^a_b} \tensor{U}{_{(3)}^c_d}(w))}{(z-w)^2}+\frac{\tensor{\delta}{^c_b} \tensor{U}{_{(4)}^a_d}(w)-\tensor{\delta}{^a_d} \tensor{U}{_{(4)}^c_b}(w)}{z-w}\ , \\
\tensor{U}{_{(2)}^a_b}&(z)\tensor{U}{_{(3)}^c_d}(w)\nonumber\\
 &\sim\frac{(n-2) (n-1) n \kappa  \left(\kappa(3 n \kappa^2-\kappa^2-2) \tensor{\delta}{^a_d} \tensor{\delta}{^c_b} +(3 n \kappa^2-2\kappa^2-1)\tensor{\delta}{^a_b} \tensor{\delta}{^c_d}\right)}{(z-w)^5}\nonumber\\
&\qquad+\frac{(n-2) (n-1)^2 \kappa ^2\tensor{\delta}{^c_d} \tensor{U}{_{(1)}^a_b}(w) +(n-2) (n-1) \left(n \kappa ^2-1\right) \kappa\tensor{\delta}{^c_b} \tensor{U}{_{(1)}^a_d}(w)  }{(z-w)^4}\nonumber\\
&\qquad+\frac{-2 (n-2) (n-1) \left(n \kappa ^2-1\right) \tensor{\delta}{^a_d} \tensor{U}{_{(1)}^c_b}(w) \kappa}{(z-w)^4}\nonumber\\
&\qquad-\frac{ (n-2) (n-1) \left(4 n \kappa ^2-3 \kappa ^2-1\right) \tensor{\delta}{^a_b} \tensor{U}{_{(1)}^c_d}(w)}{2(z-w)^4}\nonumber\\
&\qquad+\frac{(n-2) (n-1) \kappa}{(z-w)^3}  \big(n \kappa ^2\tensor{\delta}{^c_b} \partial\tensor{U}{_{(1)}^a_d}(w) -\kappa(\tensor{U}{_{(1)}^a_d}\tensor{U}{_{(1)}^c_b})(w)  \nonumber\\
&\qquad\qquad\qquad+n\kappa \tensor{\delta}{^c_d} \partial\tensor{U}{_{(1)}^a_b}(w)  -(\tensor{U}{_{(1)}^a_b}\tensor{U}{_{(1)}^c_d})(w)\big)\nonumber\\
&\qquad+\frac{(n-2) \kappa  \tensor{\delta}{^c_d} \tensor{U}{_{(2)}^a_b}(w)+(n-2) \tensor{\delta}{^c_b} \tensor{U}{_{(2)}^a_d}(w)}{(z-w)^3}\nonumber\\
&\qquad+\frac{(n-2) \left(n \kappa ^2+\kappa ^2-1\right) \tensor{\delta}{^a_d} \tensor{U}{_{(2)}^c_b}(w)+(n-2) n \kappa  \tensor{\delta}{^a_b} \tensor{U}{_{(2)}^c_d}(w)}{(z-w)^3}\nonumber\\
&\qquad+\frac{(n-2)}{2(z-w)^2} \Big(n^2\kappa ^3 \tensor{\delta}{^c_b} \partial^2\tensor{U}{_{(1)}^a_d}(w) -n\kappa ^3 \tensor{\delta}{^c_b} \partial^2\tensor{U}{_{(1)}^a_d}(w) \nonumber\\
&\qquad\qquad\qquad-2 n  \kappa ^2\left(\partial\tensor{U}{_{(1)}^a_d}\tensor{U}{_{(1)}^c_b}\right)(w)+2 \kappa ^2\left(\partial\tensor{U}{_{(1)}^a_d}\tensor{U}{_{(1)}^c_b}\right)(w) \nonumber\\
&\qquad\qquad\qquad+n^2 \kappa ^2\tensor{\delta}{^c_d} \partial^2\tensor{U}{_{(1)}^a_b}(w) -n \kappa ^2\tensor{\delta}{^c_d} \partial^2\tensor{U}{_{(1)}^a_b}(w) \nonumber\\
&\qquad\qquad\qquad-2 n \kappa \left(\partial\tensor{U}{_{(1)}^a_b}\tensor{U}{_{(1)}^c_d}\right)(w)  +2\kappa \left(\partial\tensor{U}{_{(1)}^a_b}\tensor{U}{_{(1)}^c_d}\right)(w)  \nonumber\\
&\qquad\qquad\qquad+2 \kappa(\tensor{U}{_{(1)}^a_d}\tensor{U}{_{(2)}^c_b})(w)  +2 (\tensor{U}{_{(1)}^a_b}\tensor{U}{_{(2)}^c_d})(w)\Big)\nonumber\\
&\qquad-\frac{\tensor{\delta}{^c_d} \tensor{U}{_{(3)}^a_b}(w)+\kappa  \tensor{\delta}{^c_b} \tensor{U}{_{(3)}^a_d}(w)+2 \kappa  \tensor{\delta}{^a_d} \tensor{U}{_{(3)}^c_b}(w)+2 \tensor{\delta}{^a_b} \tensor{U}{_{(3)}^c_d}(w)}{(z-w)^2}\nonumber\\
&\qquad+\frac{-2 \left(\partial\tensor{U}{_{(1)}^a_b}\tensor{U}{_{(2)}^c_d}\right)(w)+n \left(\partial\tensor{U}{_{(1)}^a_b}\tensor{U}{_{(2)}^c_d}\right)(w)}{z-w}\nonumber\\
&\qquad+\frac{-2 \kappa  \left(\partial\tensor{U}{_{(1)}^a_d}\tensor{U}{_{(2)}^c_b}\right)(w)+n \kappa  \left(\partial\tensor{U}{_{(1)}^a_d}\tensor{U}{_{(2)}^c_b}\right)(w)}{z-w}\nonumber\\
&\qquad+\frac{-\kappa  \left(\partial^2\tensor{U}{_{(1)}^a_b}\tensor{U}{_{(1)}^c_d}\right)(w)+\frac{3}{2} n \kappa  \left(\partial^2\tensor{U}{_{(1)}^a_b}\tensor{U}{_{(1)}^c_d}\right)(w)}{z-w}\nonumber\\
&\qquad+\frac{-\frac{1}{2} n^2 \kappa  \left(\partial^2\tensor{U}{_{(1)}^a_b}\tensor{U}{_{(1)}^c_d}\right)(w)-\kappa ^2 \left(\partial^2\tensor{U}{_{(1)}^a_d}\tensor{U}{_{(1)}^c_b}\right)(w)}{z-w}\nonumber\\
&\qquad+\frac{\frac{3}{2} n \kappa ^2 \left(\partial^2\tensor{U}{_{(1)}^a_d}\tensor{U}{_{(1)}^c_b}\right)(w)-\frac{1}{2} n^2 \kappa ^2 \left(\partial^2\tensor{U}{_{(1)}^a_d}\tensor{U}{_{(1)}^c_b}\right)(w)}{z-w}\nonumber\\
&\qquad+\frac{-(\tensor{U}{_{(1)}^a_d}\tensor{U}{_{(3)}^c_b})(w)+(\tensor{U}{_{(1)}^c_b}\tensor{U}{_{(3)}^a_d})(w)}{z-w}\nonumber\\
&\qquad+\frac{-\kappa  \tensor{\delta}{^c_b} \partial\tensor{U}{_{(3)}^a_d}(w)-\tensor{\delta}{^a_b} \partial\tensor{U}{_{(3)}^c_d}(w)}{z-w}\nonumber\\
&\qquad+\frac{\frac{1}{6} (n-2) (n-1) n \kappa^2(\kappa\tensor{\delta}{^c_b} \partial^3\tensor{U}{_{(1)}^a_d}(w) + \tensor{\delta}{^c_d} \partial^3\tensor{U}{_{(1)}^a_b}(w))}{z-w}\nonumber\\
&\qquad+\frac{\tensor{\delta}{^c_b} \tensor{U}{_{(4)}^a_d}(w)-\tensor{\delta}{^a_d} \tensor{U}{_{(4)}^c_b}(w)}{z-w}\ .
\end{align}
\section{Character analysis} \label{app:characters}
In this appendix, we discuss the character of the matrix-extended $\mathcal{W}_{1+\infty}$ algebra, as well as the decomposition into the sub $\mathcal{W}_{1+\infty}$ algebras. This analysis is similar to what appeared in \cite{Gaberdiel:2017hcn,Prochazka:2017qum}.
\subsection{The full vacuum character}
The matrix-extended algebra contains $m^2$ higher spin fields of every integer spin. Hence, its (generic) vacuum character is given by
\be 
\chi_\text{vac}(z_1,\dots,z_m;\tau)=\prod_{s=1}^\infty \prod_{n=s}^\infty \prod_{i,j=1}^m \frac{1}{1-y_i y_j^{-1} q^n}=\prod_{n=1}^\infty \prod_{i,j=1}^m \frac{1}{(1-y_i y_j^{-1} q^n)^n}\ ,
\ee
where $y_i=\mathrm{e}^{2\pi i z_i}$ are chemical potentials associated to the $m$ commuting $\widehat{\mathfrak{u}}(1)$ currents in the algebra, and $q=\mathrm{e}^{2\pi i \tau}$ as usual. This is a flavored version of the MacMahon function.
\subsection{Characters of \texorpdfstring{$\mathcal{W}_{1+\infty}$}{W1infty}}
\begin{figure}
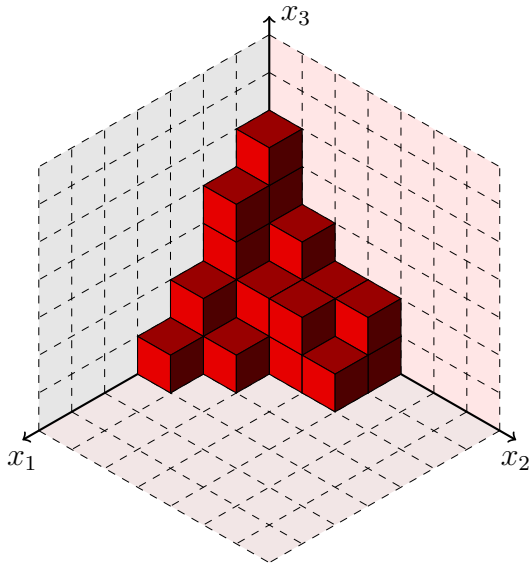

\begin{center}
\planepartition{{5,3,2,2},{4,2,2,1},{2,1},{1}}
\end{center}
\caption{Example state in the vacuum representation. This state has conformal weight $h=25$, which equals the number of boxes.} \label{fig:example state vacuum representation}
\end{figure}

Next, we recall some simple facts about $\mathcal{W}_{1+\infty}$ characters. They are best thought of in terms of plane partitions, using the isomorphism of $\mathcal{W}_{1+\infty}$ with the affine Yangian \cite{Prochazka:2015deb, Datta:2016cmw, Prochazka:2017qum, Gaberdiel:2017dbk, Gaberdiel:2017hcn, Gaberdiel:2018nbs, Prochazka:2018tlo}. In the Yangian picture, characters can be read off from counting certain plane partitions. For instance, the vacuum character is simply given by all the plane partitions with trivial asymptotics along the three axes. An example state of this form is displayed in Figure~\ref{fig:example state vacuum representation}.
Non-trivial representations can have asymptotics along the three coordinate axes, which are captured by Young diagrams. Conjugation of $\mathcal{W}_{1+\infty}$ forces one to consider also representations involving anti-boxes. Thus, the asymptotics along every axis is specified by a pair of Young diagrams $(\mu,\overline{\mu})$ of boxes and anti-boxes, respectively.

We shall be particularly interested in representations, which have only non-trivial asymptotics along one of the coordinate axes, which we conventionally take to be the $x_1$-axis. An example state in the representation with asymptotics $(\,{\tiny \yng(2,1)},\,\overline{\bullet})$ along the $x_1$-axis is displayed in Figure~\ref{fig:example state 2,1 representation}.
\begin{figure}
\begin{center}
\planepartition{{5,3,2,2},{4,2,2,1},{2,2},{2,1},{2,1},{2,1},{2,1}}
\end{center}
\caption{Example state in the representation $\left(\left({\tiny \protect\yng(2,1)},\overline{\bullet}\right),\left(\bullet,\overline{\bullet}\right),\left(\bullet,\overline{\bullet}\right)\right)$ (where $\bullet$ means trivial asymptotics). This state has conformal weight $h=16+\tfrac{3}{2}\lambda_1+\tfrac{5}{2}$, see \eqref{eq:2,1 wedge character}.}\label{fig:example state 2,1 representation}
\end{figure}
Characters with this type of asymptotics are particularly easy to determine \cite{Prochazka:2015deb,Gaberdiel:2015wpo} and it is useful to factor the character into a vacuum contribution and a `wedge'-character:
\be 
\chi_{((\mu,\overline{\mu}),(\bullet,\overline{\bullet}),(\bullet,\overline{\bullet}))}(\tau)=\chi^\text{wedge}_\mu(\tau)\chi^\text{wedge}_{\overline{\mu}}(\tau)\prod_{n=1}^\infty \frac{1}{(1-q^n)^n}\ ,
\ee
where $\chi^\text{wedge}_\mu(\tau)$ is given by
\be 
\chi^\text{wedge}_\mu(\tau)=q^{h_{1+\infty}(\mu)}\prod_{(i,j) \in \mu} \frac{1}{1-q^{h(i,j)}}\ .
\ee
Here, $(i,j)$ runs over all boxes in the Young diagrams $\mu$ and $h(i,j)$ denotes the corresponding hook length. The ground state conformal weight is given by \cite{Prochazka:2015deb}
\be 
h_{1+\infty}(\mu)=-\frac{\lambda_1}{2\lambda_2}\sum_j \mu_j^2-\frac{\lambda_1}{2\lambda_3}\sum_j (\mu \tran)^2_j+\frac{\lambda_1}{2}\sum_j \mu_j\ .
\ee
We denoted by $\mu_j$ the height of the $j$-th column and by $(\mu\tran)_j$ the length of the $j$-th row. We also made use of the $\lambda$-parameters in \eqref{eq:lambdankappa translation}.
The same formulae hold for anti-boxes and the contribution to the conformal dimensions of boxes and anti-boxes add.
 For instance, for the diagram $\left(\,{\tiny \yng(2,1)},\overline{\bullet}\right)$, we have
\be 
\chi^\text{wedge}_{\left(\,{\tiny \yng(2,1)},\,\overline{\bullet}\right)}(\tau)=\frac{q^{\frac{3}{2}\lambda_1+\frac{5}{2}}}{(1-q)^2(1-q^3)}\ . \label{eq:2,1 wedge character}
\ee
\subsection{Gluing fields}
We observe from \eqref{eq:lambda ell parameters} that $\lambda^\ell_2=-\lambda^{\ell+1}_1$, which means that this is the direction in which the two sub $\mathcal{W}_{1+\infty}$ algebras are glued together. In the following, we focus again mostly on the case of $m=2$, where there is only one gluing. In this case, matter fields sit in representations $((\bullet,\overline{\bullet}),(\mu,\overline{\mu}),(\bullet,\overline{\bullet}))$ and $((\nu,\overline{\nu}),(\bullet,\overline{\bullet}),(\bullet,\overline{\bullet}))$ with respect to the two $\mathcal{W}_{1+\infty}$ subalgebras. We have the identity
\be 
\prod_{n=1}^\infty \frac{1}{(1-q^n)^{2n}}=\sum_{\mu,\,\overline{\mu}} \chi_\mu^\text{wedge}(\tau) \chi_{\overline{\mu}}^\text{wedge}(\tau)\  \chi_{\overline{\mu}}^\text{wedge}(\tau)\chi_\mu^\text{wedge}(\tau)\ ,
\ee
where the first two wedge characters belong to the first $\mathcal{W}_{1+\infty}$ algebra and the second two wedge characters are wedge characters of the second $\mathcal{W}_{1+\infty}$ algebra. Thus, the gluing matter transforms in the representation\footnote{Note that we conjugated the representation of the second $\mathcal{W}_{1+\infty}$ with respect to the first.  This operation is invisible for the character, but is forced to be so since the gluing matter should be oppositely charged with respect to the two $\mathfrak{u}(1)$'s. }
\be 
\bigoplus_{\mu,\, \overline{\mu}} \ ((\bullet,\overline{\bullet}),(\mu,\overline{\mu}),(\bullet,\overline{\bullet}))\otimes ((\overline{\mu},\mu),(\bullet,\overline{\bullet}),(\bullet,\overline{\bullet}))
\ee
with respect to the two $\mathcal{W}_{1+\infty}$ subalgebras.
Note that this is essentially identical to \cite{Gaberdiel:2017hcn}, except that the second Young diagram is not transposed. Thus, the gluing fields in the matrix-extended algebra run over all representations with non-trivial same asymptotics in direction 2 of the first $\mathcal{W}_{1+\infty}$ algebra and direction 1 of the second algebra.

\section{Null-vectors for higher \texorpdfstring{$\boldsymbol{m}$}{m}} \label{app:null-vectors higher m}
In this appendix, we collect our data for the null-vectors for higher $m$ in the matrix-extended algebra. 

\renewcommand{\arraystretch}{4.7}
\begin{longtable}[c]{c|cccc}
Level 2 & \gluingthree{2}{1}{0}{0}{0}{0}{$n \kappa-1$}{[2,2]} & \gluingthree{2}{0}{3}{2}{1}{0}{$n -1$}{[1,1]} & \gluingthree{2}{0}{0}{1}{2}{3}{$n \kappa+\kappa+3$}{[1,1]} & \gluingthree{2}{0}{1}{1}{1}{1}{$n \kappa+1$}{[1,1]} \\
\hline
Level 3 & \gluingthree{3}{2}{0}{0}{0}{0}{$n \kappa-2$}{[3,3]} & \gluingthree{3}{0}{6}{4}{2}{0}{$n -2$}{[1,1]} & \gluingthree{3}{0}{0}{2}{4}{6}{$n \kappa+2\kappa+6$}{[1,1]}  & \gluingthree{3}{0}{2}{2}{2}{2}{$n \kappa+2$}{[0,0]} \\
\hline
\multirow{2}{*}{\vspace{-1cm} Level 4} & \gluingthree{4}{3}{0}{0}{0}{0}{$n \kappa-3$}{[4,4]} & \gluingthree{4}{1}{3}{2}{1}{0}{$n\kappa-\kappa-1$}{[2,2]} & \gluingthree{4}{1}{0}{1}{2}{3}{$n\kappa+\kappa+2$}{[2,2]} & \gluingthree{4}{0}{9}{6}{3}{0}{$n-3$}{[1,1]}  \\
\nopagebreak
 & \gluingthree{4}{0}{0}{3}{6}{9}{$n \kappa+3\kappa+9$}{[1,1]} & \gluingthree{4}{0}{1}{2}{3}{4}{$n\kappa+\kappa+4$}{[1,1]} & \gluingthree{4}{0}{4}{3}{2}{1}{$n\kappa-\kappa+1$}{[1,1]} \\
\hline
Level 5 & \gluingthree{5}{4}{0}{0}{0}{0}{$n \kappa-4$}{[5,5]}  & \gluingthree{5}{0}{12}{8}{4}{0}{$n-4$}{[1,1]}  & \gluingthree{5}{0}{0}{4}{8}{12}{$n\kappa+4\kappa+12$}{[1,1]} \\
\hline
\multirow{3}{*}{\vspace{-2cm} Level 6} &\gluingthree{6}{5}{0}{0}{0}{0}{$n \kappa-5$}{[6,6]} & \gluingthree{6}{2}{3}{2}{1}{0}{$n \kappa-\kappa-2$}{[3,3]} & \gluingthree{6}{2}{0}{1}{2}{3}{$n\kappa+\kappa+1$}{[3,3]} & \gluingthree{6}{1}{6}{4}{2}{0}{$n \kappa-2\kappa-1$}{[2,2]}  \\
\nopagebreak
& \gluingthree{6}{1}{0}{2}{4}{6}{$n \kappa+2\kappa+5$}{[2,2]}&\gluingthree{6}{0}{7}{5}{3}{1}{$n \kappa-2\kappa+1$}{[1,1]} &\gluingthree{6}{0}{1}{3}{5}{7}{$n \kappa+2\kappa+7$}{[1,1]}& \gluingthree{6}{0}{15}{10}{5}{0}{$n-5$}{[1,1]} \\
\nopagebreak
& \gluingthree{6}{0}{0}{5}{10}{15}{$n \kappa+5\kappa+15$}{[1,1]} & \gluingthree{6}{0}{3}{3}{3}{3}{$n \kappa+3$}{[1,1]} &\gluingthree{6}{0}{5}{4}{3}{2}{$n \kappa+\kappa+5$}{[0,0]}& \gluingthree{6}{0}{2}{3}{4}{5}{$n \kappa-\kappa+2$}{[0,0]} \vspace{-.5cm}\\
\caption{Truncations for $m=3$. } \label{tab:truncations m=3}
\end{longtable}
\renewcommand{\arraystretch}{1}

\renewcommand{\arraystretch}{5.4}
\begin{longtable}[c]{c|cccc}
Level 2 & \gluingfour{2}{1}{0}{0}{0}{0}{0}{$n \kappa-1$}{[2,0,2]} & \gluingfour{2}{0}{4}{3}{2}{1}{0}{$n -1$}{[1,0,1]} & \gluingfour{2}{0}{0}{1}{2}{3}{4}{$n \kappa+\kappa+4$}{[1,0,1]} & \gluingfour{2}{0}{1}{1}{1}{1}{1}{$n \kappa+1$}{[1,0,1]} \\
\hline
Level 3 & \gluingfour{3}{2}{0}{0}{0}{0}{0}{$n \kappa-2$}{[3,0,3]} & \gluingfour{3}{0}{8}{6}{4}{2}{0}{$n -2$}{[1,0,1]} & \gluingfour{3}{0}{0}{2}{4}{6}{8}{$n \kappa+2\kappa+8$}{[1,0,1]}  & \gluingfour{3}{0}{2}{2}{2}{2}{2}{$n \kappa+2$}{[1,0,1]} \\
\hline
\multirow{2}{*}{\vspace{-1cm} Level 4} & \gluingfour{4}{3}{0}{0}{0}{0}{0}{$n \kappa-3$}{[4,0,4]} & \gluingfour{4}{1}{4}{3}{2}{1}{0}{$n\kappa-\kappa-1$}{[2,0,2]} & \gluingfour{4}{1}{0}{1}{2}{3}{4}{$n\kappa+\kappa+3$}{[2,0,2]} & \gluingfour{4}{0}{12}{9}{6}{3}{0}{$n-4$}{[1,0,1]}  \\
\nopagebreak
 & \gluingfour{4}{0}{0}{3}{6}{9}{12}{$n \kappa+3\kappa+12$}{[1,0,1]} & \gluingfour{4}{0}{1}{2}{3}{4}{5}{$n\kappa+\kappa+5$}{[1,0,1]} & \gluingfour{4}{0}{5}{4}{3}{2}{1}{$n\kappa-\kappa+1$}{[1,0,1]}& \gluingfour{4}{0}{3}{3}{3}{3}{3}{$n\kappa+3$}{[0,0,0]}  \\
\hline
Level 5 & \gluingfour{5}{4}{0}{0}{0}{0}{0}{$n \kappa-4$}{[5,0,5]}  & \gluingfour{5}{0}{16}{12}{8}{4}{0}{$n-4$}{[1,0,1]}  & \gluingfour{5}{0}{0}{4}{8}{12}{16}{$n\kappa+4\kappa+16$}{[1,0,1]} \\
\hline 
\multirow{3}{*}{\vspace{-2cm} Level 6} &\gluingfour{6}{5}{0}{0}{0}{0}{0}{$n \kappa-5$}{[6,0,6]} & \gluingfour{6}{2}{4}{3}{2}{1}{0}{$n \kappa-\kappa-2$}{[3,0,3]} & \gluingfour{6}{2}{0}{1}{2}{3}{4}{$n\kappa+\kappa+2$}{[3,0,3]} & \gluingfour{6}{1}{8}{6}{4}{2}{0}{$n \kappa-2\kappa-1$}{[2,0,2]}  \\
\nopagebreak
& \gluingfour{6}{1}{0}{2}{4}{6}{8}{$n \kappa+2\kappa+7$}{[2,0,2]} & \gluingfour{6}{0}{9}{7}{5}{3}{1}{$n \kappa-2\kappa+1$}{[1,0,1]}& \gluingfour{6}{0}{1}{3}{5}{7}{9}{$n \kappa+2\kappa+9$}{[1,0,1]} & \gluingfour{6}{0}{20}{15}{10}{5}{0}{$n-5$}{[1,0,1]}\\
\nopagebreak
& \gluingfour{6}{0}{0}{5}{10}{15}{20}{$n\kappa+5\kappa+20$}{[1,0,1]} & \gluingfour{6}{0}{6}{5}{4}{3}{2}{$n\kappa-\kappa+2$}{[1,0,1]} & \gluingfour{6}{0}{2}{3}{4}{5}{6}{$n\kappa+\kappa+6$}{[1,0,1]} & \vspace{-.5cm}\\
\caption{Truncations for $m=4$. } \label{tab:truncations m=4}
\end{longtable}
\renewcommand{\arraystretch}{1}

\renewcommand{\arraystretch}{6.2}
\begin{longtable}[c]{c|cccc}
Level 2 & \gluingfive{2}{1}{0}{0}{0}{0}{0}{$n \kappa-1$}{[2,0,0,2]} & \gluingfive{2}{0}{5}{4}{3}{2}{1}{$n -1$}{[1,0,0,1]} & \gluingfive{2}{0}{0}{1}{2}{3}{4}{$n \kappa+\kappa+5$}{[1,0,0,1]} & \gluingfive{2}{0}{1}{1}{1}{1}{1}{$n \kappa+1$}{[1,0,0,1]} \\
\hline
Level 3 & \gluingfive{3}{2}{0}{0}{0}{0}{0}{$n \kappa-2$}{[3,0,0,3]} & \gluingfive{3}{0}{8}{6}{4}{2}{0}{$n -2$}{[1,0,0,1]} & \gluingfive{3}{0}{0}{2}{4}{6}{8}{$n \kappa+2\kappa+10$}{[1,0,0,1]}  & \gluingfive{3}{0}{2}{2}{2}{2}{2}{$n \kappa+2$}{[1,0,0,1]} \\
\hline
\multirow{2}{*}{\vspace{-2cm} Level 4} & \gluingfive{4}{3}{0}{0}{0}{0}{0}{$n \kappa-3$}{[4,0,0,4]} & \gluingfive{4}{1}{5}{4}{3}{2}{1}{$n\kappa-\kappa-1$}{[2,0,0,2]} & \gluingfive{4}{1}{0}{1}{2}{3}{4}{$n\kappa+\kappa+4$}{[2,0,0,2]} & \gluingfive{4}{0}{15}{12}{9}{6}{3}{$n-4$}{[1,0,0,1]}  \\
\nopagebreak
 & \gluingfive{4}{0}{0}{3}{6}{9}{12}{$n \kappa+3\kappa+15$}{[1,0,0,1]} & \gluingfive{4}{0}{1}{2}{3}{4}{5}{$n\kappa+\kappa+6$}{[1,0,0,1]} & \gluingfive{4}{0}{6}{5}{4}{3}{2}{$n\kappa-\kappa+1$}{[1,0,0,1]}& \gluingfive{4}{0}{3}{3}{3}{3}{3}{$n\kappa+3$}{[1,0,0,1]}  \\
 \hline 
 Level 5 & \gluingfive{5}{4}{0}{0}{0}{0}{0}{$n \kappa-4$}{[5,0,0,5]}  & \gluingfive{5}{0}{20}{16}{12}{8}{4}{$n-4$}{[1,0,0,1]}  & \gluingfive{5}{0}{0}{4}{8}{12}{16}{$n\kappa+4\kappa+20$}{[1,0,0,1]} & \gluingfive{5}{0}{4}{4}{4}{4}{4}{$n\kappa+4$}{[0,0,0,0]}\\
\caption{Truncations for $m=5$. } \label{tab:truncations m=5}
\end{longtable}
\renewcommand{\arraystretch}{1}

\clearpage

\bibliographystyle{JHEP}
\bibliography{bib}

\end{document}